\documentclass[aps, prl, twocolumn, superscriptaddress, floatfix]{revtex4-2}
\usepackage{graphicx}% Include figure files
\usepackage{bm, amsmath,amssymb}% bold math
\usepackage{float}
\usepackage{color}
\usepackage{multirow}
\usepackage{newtxtext,newtxmath}
\usepackage{hyperref}
\hypersetup{hypertex=true, colorlinks=true, linkcolor=blue, anchorcolor=blue,citecolor=blue, urlcolor=blue}

\begin{document}

\title{Doniach phase diagram for Kondo lattice model on the square and triangular lattices}

\author{Ruixiang Zhou}
\affiliation{School of Physical Science and Technology, ShanghaiTech University, Shanghai 201210, China}
\author{ Xuefeng Zhang}
\affiliation{School of Physical Science and Technology, ShanghaiTech University, Shanghai 201210, China}
\author{Gang Li}
\email{ligang@shanghaitech.edu.cn}
\affiliation{School of Physical Science and Technology, ShanghaiTech University, Shanghai 201210, China}
\affiliation{\mbox{ShanghaiTech Laboratory for Topological Physics, ShanghaiTech University, Shanghai 201210, China}}

\date{\today}

\begin{abstract}
Geometric frustration adds a new competing energy scale to the antiferromagnetic (AFM) Kondo lattice model (KLM).
In this work, we systematically study the doniach phase diagram on the square and triangular lattices in the same theoretical framework and reveal unexpected responses of it on the two lattices.
The potential energy created by the geometric frustration is comparable to the Ruderman-Kittel-Kasuya-Yosida (RKKY) coupling, which completely suppresses the long-range antiferromagnetic (AFM) order on the half-filled triangular lattice. While, on the square lattice, the long-range AFM order successfully establishes and constitutes the conventional competition between the RKKY and Kondo couplings. 
The geometrical frustration on the triangular lattice is partially released when doped with holes, in which two different magnetic orders emerge unexpectedly. 
The two orders closely relate to the topology of the interacting Fermi surface. 
Our comprehensive comparison of the KLM on the two lattices reveals a significant competition of geometric frustration, RKKY, and Kondo couplings on low-dimensional systems and sheds light on experimentally finding new phases in related materials. 
\end{abstract}

\maketitle

{\it Introduction --.}
Heavy fermion compounds are strongly correlated electron systems with an extraordinarily large effective electron mass, as experimentally reflected in the large linear $T$-dependence of the specific heat, i.e., $C_V=\gamma*T$, as well as the enhanced Pauli susceptibility $\chi(T)$ at temperatures below the so-called Kondo temperature $T_{K}$~\cite{PhysRevLett.35.1779, FULDE19881, onuki1987heavy, steglich1985heavy, RevModPhys.56.755}. 
An accepted low-energy description of the heavy fermion systems is the Kondo lattice model (KLM)~\cite{coleman2015introduction}.
In this model, conduction electrons scatter from local moments with coupling strength $J$. 
There are two fundamental energy scales in this model. 
One is the superexchange coupling between local moments mediated by the conduction electrons, which was first considered by M. Ruderman and C. Kittel~\cite{PhysRev.96.99}, then further elaborated by T. Kasuya~\cite{1956PThPh1645K} and K. Yosida~\cite{PhysRev.106.893}, which is now known as the RKKY interaction $K\propto J^{2}$. 
On the other hand, when $J$ is large, driven by the antiferromagnetic coupling $J$, some conduction electrons form a spin singlet with the local moment and lose their mobility below $T_{K}$. 
The local moment is screened by the spin of these conduction electrons as if it was effectively removed from the system. 
This process is known as the Kondo screening. 
Other electrons will no longer experience the presence of the local moment but only scatter over an effective potential, leading to a Fermi liquid behavior of them~\cite{Pepin_Kondo, Gulacsi2006-GULTKL}.   

The RKKY interaction and the Kondo screening dominate in the small and large Kondo coupling $J$ region and strongly compete in the intermediate regime, qualitatively described by doniach phase diagram~\cite{DONIACH1977231, Doniach1977}. 
In this letter, we want to carefully compare the doniach phase diagram on the square and triangular lattices with an advanced many-body algorithm, i.e. the dual-fermion (DF) approach~\cite{PhysRevB.79.045133, PhysRevLett.102.206401}. It is a non-local extension of the dynamical mean-field theory (DMFT)~~\cite{georges1996dynamical, burdin2002heavy, otsuki2009evolution, peters2013charge}  and has been verified against numerically exact methods~\cite{PhysRevX.11.011058}. (See the Supplementary Information for more details). 
The different lattice geometries add an additional ingredient to the conduction electrons, which itself can mediate their magnetic couplings. 
What is the qualitative difference between the RKKY interactions and Kondo screening on the two different lattices is an interesting question to understand. 
Although many theoretical studies have been conducted for the KLM and other related models~\cite{fazekas1991magnetic, burdin2002heavy, otsuki2009evolution, lacroix1979phase, peters2013charge, martin2010fermi, martin2008evolution, doi:10.1143/JPSJ.78.014702, doi:10.1143/JPSJ.78.034719, akagi2012hidden, aulbach2015dynamical, kessler2020magnetic, PhysRevB.102.155126}, 
a decent comparison of the two lattices in the same theoretical framework over a wide temperature and doping regime is still lacking, from which we expect to extract a convincing conclusion on their characteristic difference.

\begin{figure*}[t]
	\centering
	\includegraphics[width=0.9\linewidth]{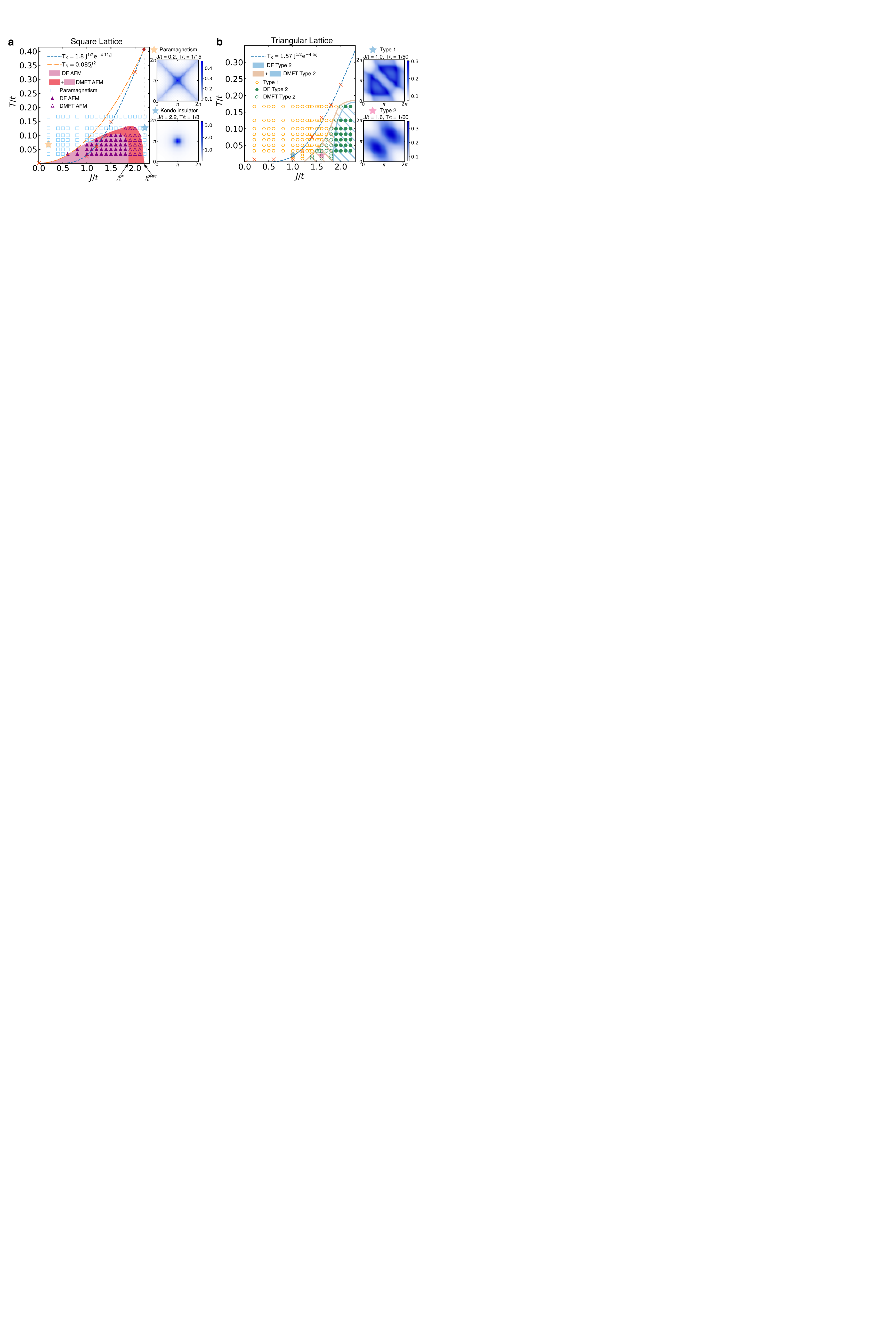}
	\caption{
\textbf{(Color Online) Doniach phase diagram for the half-filled square and triangular lattice.}
\textbf{a}. On the square lattice, the RKKY coupling induces a long-range AFM order at a small to intermediate $J$ regime highlighted in color. The pin and red colored region correspond to the AFM phase from the DF and DMFT calculations, respectively. The different symbols denote the different solutions from the two methods. $T_{N}$ and $T_{K}$ were obtained from the AFM phase boundary and the inverse DMFT uniform susceptibility.
On the right-hand side, the two spin susceptibility plots correspond to two characteristic parameters for the paramagnetic metal and Kondo insulator with $J/t=0.2/2.2$, and $T=0.05/0.10$, respectively.
\textbf{b}. On the triangular lattice,  long-range AFM order is absent. The Kondo coupling is the only dominating energy scale. Two characteristic spin susceptibility behaviors were observed and denoted as type 1 and type 2 corresponding to the yellow and green circles. Examples of the two susceptibilities are shown on the right plots with $J/t=1.2/1.6$, $T/t=0.1667$.
The regime shaded with line pattern correspond to the parameters where spin susceptibility displays a typical 120$^\circ$-AFM correlation (Type 2), but no long-range order. 
}
\label{doniach_diagram}
\end{figure*}

{\it Doniach phase diagram at half-filling --.}
The most striking difference of the KLM on the square and triangular lattice is the absence of the long-range magnetic order of the conduction electrons in the RKKY regime on the triangular lattice, see Fig.~\ref{doniach_diagram}\textbf{b}. 
On the square lattice, we observed a well-defined antiferromagnetic (AFM) long-range order with magnetic wave vector $\vec{Q}=(\pi, \pi)$ in Fig.~\ref{doniach_diagram}\textbf{a}. 
The finite temperature AFM phase is due to the approximation inherent in our many-body algorithms (see the Supplementary Information for an introduction to the methodology).
In real materials, the presence of other types of magnetic interaction, for example, the spin anisotropy~\cite{Nature_CrI3, Gong_Discovery_2017, PhysRevLett.123.047203},  can stabilize the AFM phase at finite temperature and be compatible with the Mermin-Wigner theorem~\cite{PhysRevLett.17.1133}. 
The access of the local uniform susceptibility and its fluctuations calculated in DMFT allows us to extract the Kondo temperature $T_{K}\sim 1.8J^{1/2}e^{-4.11/J}$ for the square lattice and $T_{K}\sim 1.57J^{1/2}e^{-4.5/J}$ for the triangular lattice shown as the blue dashed line in Fig.~\ref{doniach_diagram}~\cite{DONIACH1977231, PhysRevB.28.5255}.  
The crossing point of $T_{K}$ and $T_{N}$ nicely coincides with $J^{DMFT}_{c}$ estimated from the destruction of the AFM phase in DMFT on the square lattice. 

The obvious difference in the KLM on square and triangular lattices highlights the underlying geometrical influence on the conduction electrons. 
In addition to the competition of the RKKY and Kondo couplings, the geometrical frustration further competes with the two coupling strengths, destroying the long-range AFM order at the weak-$J$ regime on the triangular lattice. 
Later in the discussion for the hole-doping KLM, one will further see that, hole-doping partially releases the geometrical frustration, and the two lattices again give unexpected differences in magnetic response. 

After understanding the doniach diagram, we further discuss the MIT phase boundary, which is not shown in Fig.~\ref{doniach_diagram} but jointly determines the ground state of the KLM with the doniach phase diagram.  
The study of the MIT amounts to calculating the local density of states (DOS), which relates to the imaginary part of the single-particle Green's function as $A(\omega) = -\sum_{\vec{k}}\mbox{Im}G(\vec{k},\omega)/\pi$.
At half-filling, the KLM on both the square and triangle lattices can develop a charge gap, rising the possibility of forming both Kondo metal and Kondo insulator in the KLM. 

\begin{figure}[t]
\centering
\includegraphics[width=\linewidth]{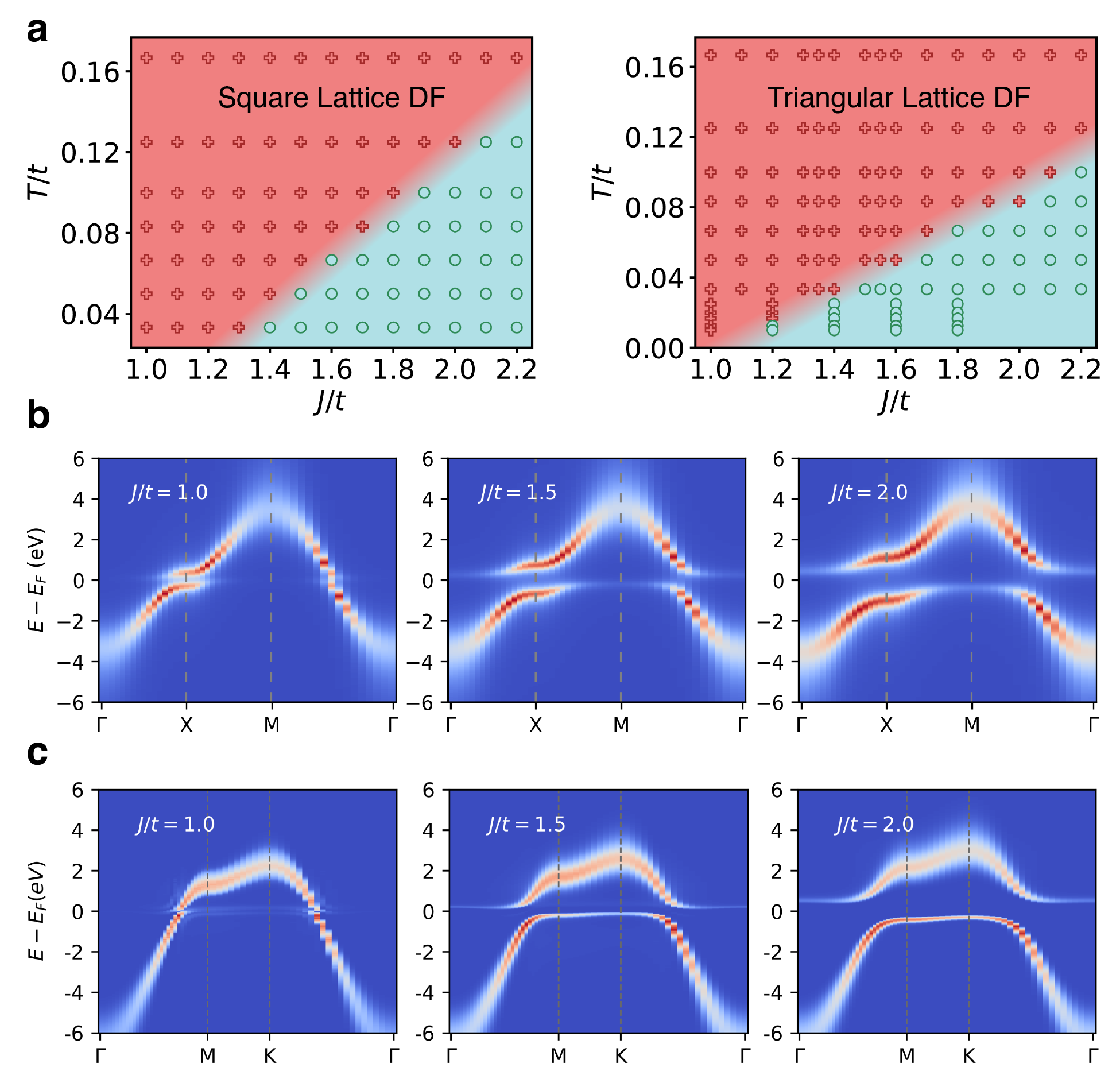}
\caption{\textbf{(Color Online) The MIT phase diagram and spectral functions of KLM on the square and triangle lattices.} 
\textbf{a} display the MIT phase boundary for the KLM on the square and triangle lattice, respectively. 
\textbf{b}. The spectral function along $\Gamma-X-M-\Gamma$ at $T/t=0.04$ and three different couplings $J/t$ on the square lattice. 
\textbf{c} shows the same results like \textbf{c}, but for the triangular lattice. 
All results in this figure are obtained with the DF method.  
}
\label{Fig_MIT}
\end{figure}

In Fig.~\ref{Fig_MIT}, the metallic and insulating solutions are shown in different symbols shaded by the light-red and light-cyan colors. 
We observe that the MIT phase boundary moves to a smaller $J_{c}$ with the decrease of temperature on both lattices, displaying a left-going phase boundary. 
On the square lattice, the MIT phase boundary is under the dome of the AFM phase in Fig.~\ref{doniach_diagram} separating it into two parts. 
At the weak $J$ regime, the conduction electrons are itinerant and AFM ordered corresponding to RKKY metals. 
For intermediate $J$, where the RKKY and Kondo couplings strongly compete, the low-temperature states are insulating that can be either RKKY-type insulators with long-range AFM order or Kondo insulators with spin singlets depending on the temperature. 
For large $J > J_{c}^{\mbox{DMFT}} \sim 2.2 > J_{c}^{\mbox{DF}}\sim 1.8$,  these states belong to Kondo insulators on the square lattice.
On the triangular lattice, the situation becomes much simpler. 
As there is no AFM long-range order on the triangular lattice, for any parameter $(J, T)$ the MIT boundary shown in Fig.~\ref{Fig_MIT}\textbf{c} and Fig.~\ref{Fig_MIT}\textbf{d} is then the phase boundary for the Kondo metal and Kondo insulator at sufficiently low temperature.

We note that the MIT transition boundary was obtained with a paramagnetic condition.
As magnetic long-range order can gap the conduction electrons, we suspect that in a spin-polarized calculation, the MIT will occur in a smaller critical $J/t$ on the square lattice~\cite{PhysRevB.81.113108}. 
However, due to the geometric frustration, no magnetic order is stabilized on the triangular lattice. 
The nature of MIT on the triangular lattice is then a paramagnetic phase transition. 
Kondo coupling becomes the only dominating energy scale on the half-filled triangular KLM. 

The presence of the AFM phase on the square lattice has stimulated an interesting question concerning the existence of the Kondo screening inside the magnetic phase. 
Our results on the spectral function shown in Fig.~\ref{Fig_MIT}\textbf{b} and \textbf{c} support this idea~\cite{PhysRevLett.99.136401, PhysRevLett.101.066404, PhysRevB.104.155128}. 
On the square lattice, at $J/t = 1.5$ where the AFM order is present, the spectral function shows a clear flat band extending from $X$ and $\Gamma$ to $M$, which is already similar to that in the Kondo insulating phase with $J/t = 2.0$.
Furthermore, on the triangular lattice, due to the weak competition of the RKKY coupling, the flat bands around the Fermi level more easily establish for the same parameter, confirming the dominating role of the Kondo coupling in the triangular lattice. 

\begin{figure}[t]
\centering
\includegraphics[width=\linewidth]{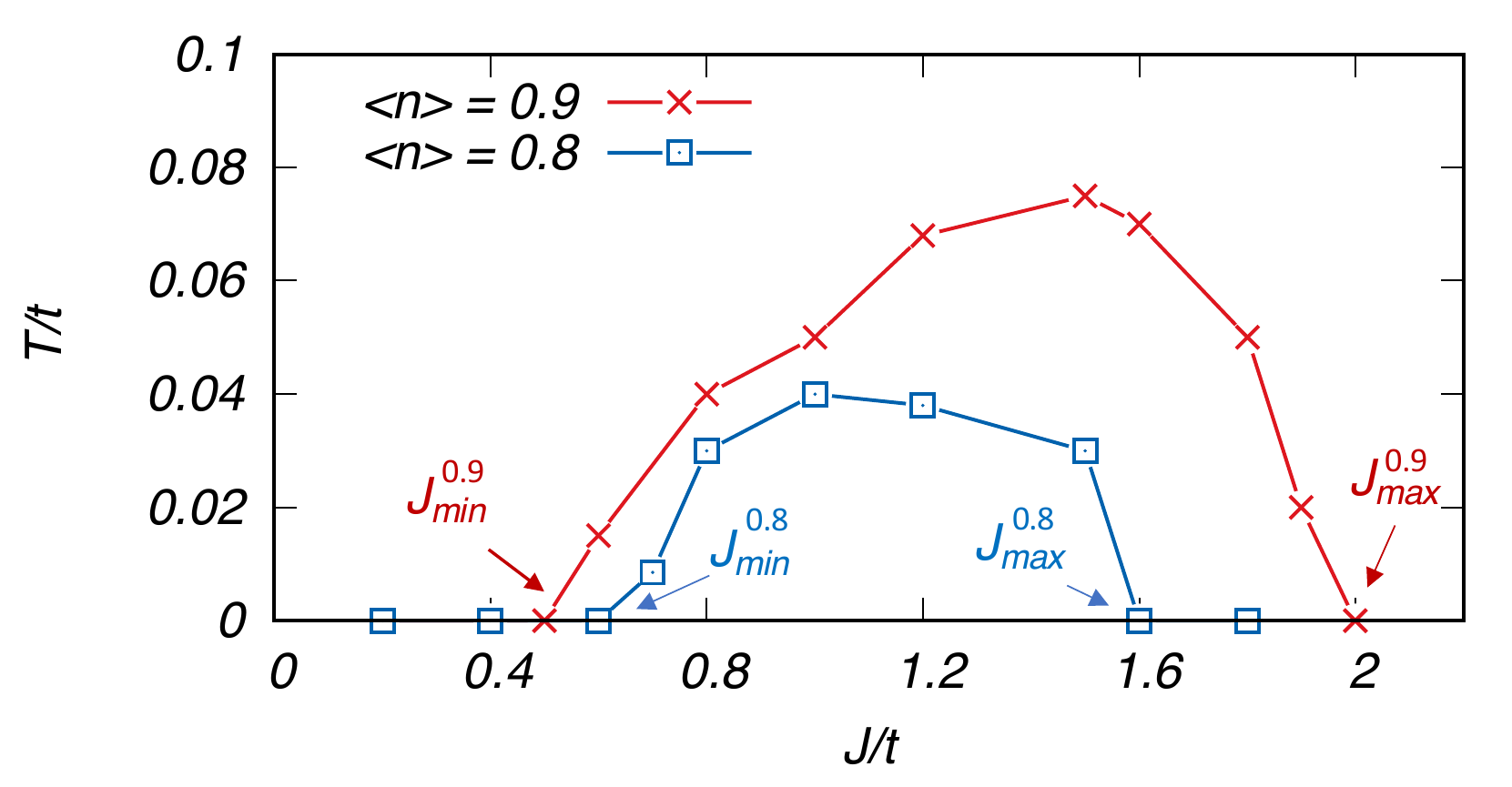}
\caption{\textbf{(Color Online) The magnetic phase diagram of the hole-doped KLM on the square lattice.}
The magnetic phase boundary for two hole-doping levels. The phase boundary is obtained from a Curie-Weiss fit of the spin susceptibility of the conduction electrons.
All results in this figure are obtained with the DF calculations.  
}
\label{hole_square}
\end{figure}

\begin{figure}[htbp]
\centering
\includegraphics[width=\linewidth]{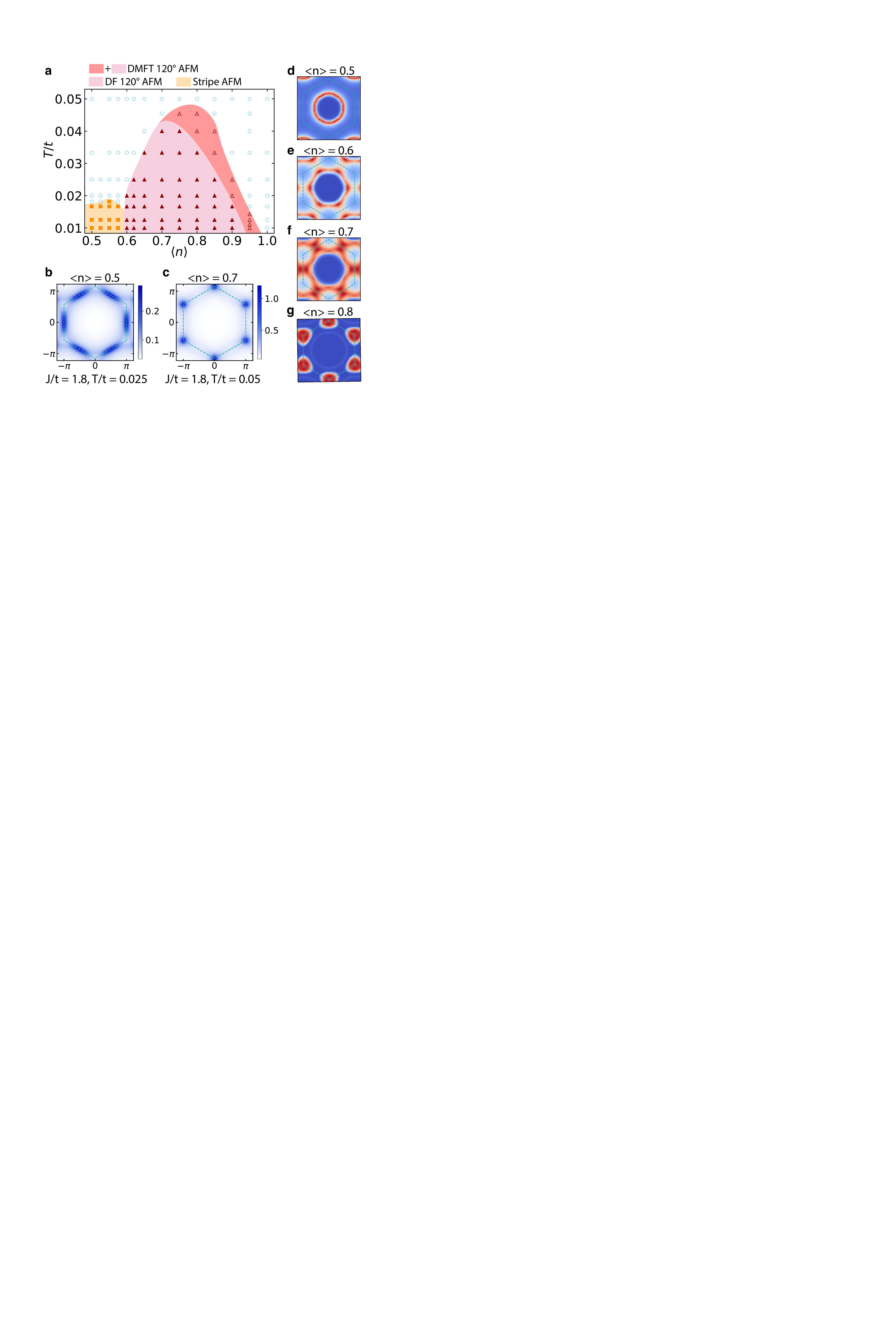}
\caption{\textbf{(Color Online) The magnetic phase diagram of the hole-doped KLM on the triangular lattice. }
\textbf{a.} There are two different magnetic orders discovered at different hole doping levels with $J/t = 1.8$. 
One is the 120$^\circ$-AFM at $0.6<\langle n\rangle < 0.9$,  and the other one is 
 row-wise AFM at $0.5<\langle n\rangle< 0.6$.
\textbf{b} and \textbf{c} illustrate the spin susceptibility at $\langle n\rangle=0.5$ and 0.7 displaying different magnetic wave vectors. 
\textbf{d-g} display the interacting Fermi surface at different electron concentrations. In $\textbf{d}$, the arrows indicate the coherent scattering wave vectors.
}
\label{hole_triangle}
\end{figure}

{\it Hole-doping --}.
\label{Sec:hole}
After understanding the half-filled case, we now move on to the hole-doped KLM model to further understand the persistence of the AFM phase and the emergence of other types of order. 
One of these phases is the unexpected magnetic long-range order on the triangular lattice. 
In the following, we will separately discuss the doped KLM model on the square and triangular lattices in Fig.~\ref{hole_square} and Fig.~\ref{hole_triangle}. 

Figure~\ref{hole_square} displays the magnetic phase diagram of the KLM on the square lattice at two different hole doping levels. 
For the doped one-band model, there are no correlation-driven or filling-driven insulating states. 
States at all parameters shown in Fig.~\ref{hole_square} are metallic. 
We observed two characteristic features of the doped KLM on the square lattice. 
The first one is the shrink of the AFM phase space with the increase of hole doping. 
With the increase of the hole-doping level, the N\`eel temperature decreases.
Our result is consistent with the observation in dynamical cluster approximation calculation~\cite{PhysRevLett.101.066404}.
Moreover, at the weak $J$ regime, the AFM phase is completely suppressed. Our finite temperature calculation does not capture any signature of long-range RKKY order for $J/t <0.5$. 
We note that $J_{max}$, where the AFM order vanishes,  becomes slightly larger for $\langle n\rangle = 0.9$ as compared to Fig.~\ref{doniach_diagram}. 
This can be an artificial effect caused by the fact that we only considered the 2nd DF self-energy diagram and did not employ a full charge self-consistency on both the DMFT and DF levels. 
Employing the ladder DF self-energy scheme~\cite{PhysRevLett.102.206401}  and the full charge self-consistency with the simultaneous convergence of the DMFT and DF solutions will likely reduce $J_{max}$ to a value smaller than that for half-filling, as observed in the single-band Hubbard model~\cite{PhysRevLett.115.036404}. 

After demonstrating the destruction of the RKKY interactions in the hole-dope KLM on the square lattice, we now turn the focus to the hole-doped triangular lattice. 
Due to the absence of long-range orders in the half-filled case, we do not expect a considerable difference with hole-doping. 
Similar to the square lattice, we anticipated that hole-doping further suppresses the RKKY coupling such that the magnetic phase diagram of the KLM on the triangular lattice would remain completely featureless. 
Instead, we found a rich magnetic phase diagram as displayed in Fig.~\ref{hole_triangle}\textbf{a}. 
We discovered two long-range magnetic orders at different doping levels. 
With the increase of hole doping, we first observe a 120$^{\circ}$-AFM for a wide range of electron occupancy $0.6<\langle n\rangle < 0.9$, shown as the pink region in Fig.~\ref{hole_triangle}. 
The N$\acute{e}$el temperature maximizes at $\langle n\rangle\sim 0.7$. 
Further increasing the hole doping, a new magnetic correlation with row-wise type spin arrangement appears when the electron occupancy is below $\langle n\rangle< 0.6$. 
The smallest electron occupancy studied in our work is $\langle n\rangle = 0.5$, which is in the row-wise AFM phase as well. 

We attribute the emergence of magnetic orders in the hole-doped triangular KLM to the partial release of geometric frustration and the Fermi surface nesting effect. 
In comparison to the square lattice, the Fermi surface is not perfectly nested at half-filling.  
There is no spontaneous spin instability for the non-interacting tight-binding model on the triangular lattice. 
In the Hubbard or periodic Anderon model, only when the coupling between two electrons is strong enough, the classical $120^{\circ}$-AFM can establish~\cite{PhysRevLett.97.046402, PhysRevB.89.161118, PhysRevB.103.235134}. 
In the KLM, the effective coupling of conduction electrons stems from the superexchange involving two scattering processes at two neighboring sites, and the coupling is proportional to $J^{2}$. It is easier for the conduction electrons to magnetically order at larger values of $J/t$, where, however, the formation of the Kondo singlet will strongly compete. 
Thus, if exists, the $120^{\circ}$-AFM will only appear at intermediate $J/t$ at half-filling. 
However, as shown in Fig.~\ref{doniach_diagram}\textbf{b}, the superexchange coupling between conduction electrons alone is not sufficient to establish a long-range order before the formation of the Kondo singlet. 
But, with hole-doping, the Fermi surface starts to play a role and triggers the emergence of two different magnetic orders. 

Away from half-filling, the interacting Fermi surface gradually changes topology. We show, in Fig.~\ref{hole_triangle}\textbf{d-g}, the Fermi surface obtained as $A(\vec{k},\omega=E_f) = -\mbox{Im}G(\vec{k},\omega)/\pi$ at different electron concentrations. 
At $\langle n\rangle=0.6$ and 0.7, the interacting Fermi surface shows a clear hexagonal shape with the nesting vectors corresponding exactly to the $\mathbf{\Gamma}-\mathbf{K}$ vector.  
Further increasing electron concentration changes the hexagon at $\mathbf{\Gamma}$ to six triangles located at $\mathbf{K}$ point. 
The strong density around $\mathbf{K}$ point supports coherent scattering between any two $\mathbf{K}$ points with the scattering vector the same as the $\mathbf{\Gamma}-\mathbf{K}$ vector. 
At $\langle n\rangle=0.9$, the size of the Fermi surface at each $\mathbf{K}$ becomes smaller leading to a reduced coherence scattering in this case. 
As a consequence, we found that N$\acute{e}$el temperature maximizes around $\langle n\rangle=0.7$ and decreases with the increase of electron concentration.
At smaller doing, for example, $\langle n\rangle=0.5$, the hexagon Fermi surface shrinks to a circle around $\mathbf{\Gamma}$ point. 
A circular Fermi surface does not have nesting vectors and usually does not trigger any magnetic instability. 
In fact, at half-filling the Fermi surface of the tight-binding model on the triangular lattice is also a circle and, as we know, no magnetic long-range order can be established. 
Here, at $\langle n\rangle=0.5$, the diameter of the interacting circular Fermi surface is the same as the $\mathbf{\Gamma}-\mathbf{M}$ vector. 
Both the intra- and inter-Fermi surface scatterings with this wave vector coherently contribute to the magnetic instability, leading to the peak structure of the spin susceptibility with the same wave vector shown in Fig.~\ref{hole_triangle}\textbf{b}.
At half-filling, however, the intra- and inter-Fermi surface scatterings correspond to different wave vectors and cannot form coherence.

{\it Discussions and Conclusions --.}
The square and triangular lattices differ in their geometry and frustration to the long-range spin arrangement. 
There have been many theoretical studies of correlated models, such as the Hubbard model~\cite{PhysRevB.40.506, PhysRevB.62.4336, PhysRevB.78.205117, PhysRevB.80.075116, PhysRevLett.72.705, PhysRevB.91.245125, PhysRevB.91.205121, PhysRevB.94.125144,PhysRevB.96.205130, doi:10.1146/annurev-conmatphys-090921-033948, PhysRevB.89.161118, PhysRevLett.100.136402, PhysRevResearch.2.013295, PhysRevB.103.235134}, on these two lattices, which establish a fruitful understanding of the different electronic and magnetic responses on them.
However, the KLM has not been thoroughly understood and compared on the two lattices. 
 
Our first important observation is the different magnetic responses of the half-filled KLM on the square and triangular lattices.  
The presence/absence of long-range AFM order on the square/triangular lattices indicates that the RKKY exchange coupling, as a superexchange coupling involving multiple Coulomb scattering processes, is sensitive and comparable in energy scale with the potential created by geometric frustrations. 
In comparison to the Hubbard model on the triangular lattice, in which the strong Coulomb repulsion is able to stabilize long-range AFM order, the KLM on the triangular lattice is featureless with the Kondo coupling being the only dominating energy scale.

However, in doping with holes, more exotic differences appear. 
The KLM on the square lattice shows the expected destruction of the AFM long-range order with the increase of hole concentration.
While on the triangular lattice, two magnetic phases appear at different hole doping levels unexpectedly. 
A 120$^{\circ}$-AFM emerges for a wide range of electron occupancy $0.6<\langle n\rangle < 0.9$.
Between $0.5 < \langle n\rangle < 0.6$, a row-wise type AFM phase with the magnetic wave vector $\vec{Q}=(\pi, 0)$ emerges. 
The analysis of the interacting Fermi surface leads to our second important observation.  
We found that the 120$^{\circ}$-AFM order is induced by the intra-Fermi-surface scattering which shows clear nesting topology and favors the coherent magnetic excitation with wave vector $\mathbf{\Gamma}-\mathbf{K}$. 
At larger hole doping, the Fermi surface topology transforms to a circular shape. 
The coherent scattering of the intra- and inter-Fermi-surface scattering promotes a row-wise AFM order. 
Due to the critical condition and the less strong coherence scatterings, the row-wise AFM phase demonstrates lower N$\acute{e}$el temperatures and smaller phase space. 

Our systematic study of the KLM on the square and triangular lattices provides strong insight into the underlying competition of the RKKY coupling and the potential energy created by the geometrical frustration. 
The doniach phase diagrams, magnetic, and electronic excitations of the KLM at both half-filling and doped cases constitute a comprehensive understanding of the KLM on low-dimensional lattices and may pave the way for the study of related materials.   

{\it Acknowledgement --.}
G.L. wants to thank the support of collaborators on other related topics, including but not limited to H. Monien, A. Rubtsov, P. Werner, F. Assaad, W. Hanke, R. Thomale, A.I. Lichenstein, and K. Held et. al. 
This work was supported by the National Key R\&D Program of China (2022YFA1402703),  National Natural Science Foundation of China (11874263), Shanghai 2021-Fundamental Research Aera (21JC1404700), Shanghai Technology Innovation Action Plan (20DZ1100605), and Sino-German Mobility program (M-0006).  
X.Z. acknowledges the Postdoctoral Special Funds for Theoretical Physics of the National Natural Science Foundation of China (12147124). 
Part of the calculations was performed at the HPC Platform of ShanghaiTech University Library and Information Services, and at the School of Physical Science and Technology.

\bibliographystyle{apsrev4-2}
\bibliography{reference.bib}

\begin{thebibliography}{62}%
\makeatletter
\providecommand \@ifxundefined [1]{%
 \@ifx{#1\undefined}
}%
\providecommand \@ifnum [1]{%
 \ifnum #1\expandafter \@firstoftwo
 \else \expandafter \@secondoftwo
 \fi
}%
\providecommand \@ifx [1]{%
 \ifx #1\expandafter \@firstoftwo
 \else \expandafter \@secondoftwo
 \fi
}%
\providecommand \natexlab [1]{#1}%
\providecommand \enquote  [1]{``#1''}%
\providecommand \bibnamefont  [1]{#1}%
\providecommand \bibfnamefont [1]{#1}%
\providecommand \citenamefont [1]{#1}%
\providecommand \href@noop [0]{\@secondoftwo}%
\providecommand \href [0]{\begingroup \@sanitize@url \@href}%
\providecommand \@href[1]{\@@startlink{#1}\@@href}%
\providecommand \@@href[1]{\endgroup#1\@@endlink}%
\providecommand \@sanitize@url [0]{\catcode `\\12\catcode `\$12\catcode
  `\&12\catcode `\#12\catcode `\^12\catcode `\_12\catcode `\%12\relax}%
\providecommand \@@startlink[1]{}%
\providecommand \@@endlink[0]{}%
\providecommand \url  [0]{\begingroup\@sanitize@url \@url }%
\providecommand \@url [1]{\endgroup\@href {#1}{\urlprefix }}%
\providecommand \urlprefix  [0]{URL }%
\providecommand \Eprint [0]{\href }%
\providecommand \doibase [0]{https://doi.org/}%
\providecommand \selectlanguage [0]{\@gobble}%
\providecommand \bibinfo  [0]{\@secondoftwo}%
\providecommand \bibfield  [0]{\@secondoftwo}%
\providecommand \translation [1]{[#1]}%
\providecommand \BibitemOpen [0]{}%
\providecommand \bibitemStop [0]{}%
\providecommand \bibitemNoStop [0]{.\EOS\space}%
\providecommand \EOS [0]{\spacefactor3000\relax}%
\providecommand \BibitemShut  [1]{\csname bibitem#1\endcsname}%
\let\auto@bib@innerbib\@empty
%</preamble>
\bibitem [{\citenamefont {Andres}\ \emph {et~al.}(1975)\citenamefont {Andres},
  \citenamefont {Graebner},\ and\ \citenamefont {Ott}}]{PhysRevLett.35.1779}%
  \BibitemOpen
  \bibfield  {author} {\bibinfo {author} {\bibfnamefont {K.}~\bibnamefont
  {Andres}}, \bibinfo {author} {\bibfnamefont {J.~E.}\ \bibnamefont
  {Graebner}},\ and\ \bibinfo {author} {\bibfnamefont {H.~R.}\ \bibnamefont
  {Ott}},\ }\href {https://doi.org/10.1103/PhysRevLett.35.1779} {\bibfield
  {journal} {\bibinfo  {journal} {Phys. Rev. Lett.}\ }\textbf {\bibinfo
  {volume} {35}},\ \bibinfo {pages} {1779} (\bibinfo {year}
  {1975})}\BibitemShut {NoStop}%
\bibitem [{\citenamefont {Fulde}\ \emph {et~al.}(1988)\citenamefont {Fulde},
  \citenamefont {Keller},\ and\ \citenamefont {Zwicknagl}}]{FULDE19881}%
  \BibitemOpen
  \bibfield  {author} {\bibinfo {author} {\bibfnamefont {P.}~\bibnamefont
  {Fulde}}, \bibinfo {author} {\bibfnamefont {J.}~\bibnamefont {Keller}},\ and\
  \bibinfo {author} {\bibfnamefont {G.}~\bibnamefont {Zwicknagl}}\ }(\bibinfo
  {publisher} {Academic Press},\ \bibinfo {year} {1988})\ pp.\ \bibinfo {pages}
  {1--150}\BibitemShut {NoStop}%
\bibitem [{\citenamefont {ŌNUKI}\ and\ \citenamefont
  {KOMATSUBARA}(1987)}]{onuki1987heavy}%
  \BibitemOpen
  \bibfield  {author} {\bibinfo {author} {\bibfnamefont {Y.}~\bibnamefont
  {ŌNUKI}}\ and\ \bibinfo {author} {\bibfnamefont {T.}~\bibnamefont
  {KOMATSUBARA}},\ }in\ \href
  {https://doi.org/https://doi.org/10.1016/B978-1-4832-2948-5.50086-1} {\emph
  {\bibinfo {booktitle} {Anomalous Rare Earths and Actinides}}},\ \bibinfo
  {editor} {edited by\ \bibinfo {editor} {\bibfnamefont {J.}~\bibnamefont
  {BOUCHERLE}}, \bibinfo {editor} {\bibfnamefont {J.}~\bibnamefont {FLOUQUET}},
  \bibinfo {editor} {\bibfnamefont {C.}~\bibnamefont {LACROIX}},\ and\ \bibinfo
  {editor} {\bibfnamefont {J.}~\bibnamefont {ROSSAT-MIGNOD}}}\ (\bibinfo
  {publisher} {Elsevier},\ \bibinfo {year} {1987})\ pp.\ \bibinfo {pages}
  {281--288}\BibitemShut {NoStop}%
\bibitem [{\citenamefont {Steglich}\ \emph {et~al.}(1985)\citenamefont
  {Steglich}, \citenamefont {Rauchschwalbe}, \citenamefont {Gottwick},
  \citenamefont {Mayer}, \citenamefont {Sparn}, \citenamefont {Grewe},
  \citenamefont {Poppe},\ and\ \citenamefont {Franse}}]{steglich1985heavy}%
  \BibitemOpen
  \bibfield  {author} {\bibinfo {author} {\bibfnamefont {F.}~\bibnamefont
  {Steglich}}, \bibinfo {author} {\bibfnamefont {U.}~\bibnamefont
  {Rauchschwalbe}}, \bibinfo {author} {\bibfnamefont {U.}~\bibnamefont
  {Gottwick}}, \bibinfo {author} {\bibfnamefont {H.~M.}\ \bibnamefont {Mayer}},
  \bibinfo {author} {\bibfnamefont {G.}~\bibnamefont {Sparn}}, \bibinfo
  {author} {\bibfnamefont {N.}~\bibnamefont {Grewe}}, \bibinfo {author}
  {\bibfnamefont {U.}~\bibnamefont {Poppe}},\ and\ \bibinfo {author}
  {\bibfnamefont {J.~J.~M.}\ \bibnamefont {Franse}},\ }\href
  {https://doi.org/10.1063/1.335212} {\bibfield  {journal} {\bibinfo  {journal}
  {Journal of Applied Physics}\ }\textbf {\bibinfo {volume} {57}},\ \bibinfo
  {pages} {3054} (\bibinfo {year} {1985})}\BibitemShut {NoStop}%
\bibitem [{\citenamefont {Stewart}(1984)}]{RevModPhys.56.755}%
  \BibitemOpen
  \bibfield  {author} {\bibinfo {author} {\bibfnamefont {G.~R.}\ \bibnamefont
  {Stewart}},\ }\href {https://doi.org/10.1103/RevModPhys.56.755} {\bibfield
  {journal} {\bibinfo  {journal} {Rev. Mod. Phys.}\ }\textbf {\bibinfo {volume}
  {56}},\ \bibinfo {pages} {755} (\bibinfo {year} {1984})}\BibitemShut
  {NoStop}%
\bibitem [{\citenamefont {Coleman}(2015)}]{coleman2015introduction}%
  \BibitemOpen
  \bibfield  {author} {\bibinfo {author} {\bibfnamefont {P.}~\bibnamefont
  {Coleman}},\ }\href@noop {} {\emph {\bibinfo {title} {Introduction to
  many-body physics}}}\ (\bibinfo  {publisher} {Cambridge University Press},\
  \bibinfo {year} {2015})\BibitemShut {NoStop}%
\bibitem [{\citenamefont {Ruderman}\ and\ \citenamefont
  {Kittel}(1954)}]{PhysRev.96.99}%
  \BibitemOpen
  \bibfield  {author} {\bibinfo {author} {\bibfnamefont {M.~A.}\ \bibnamefont
  {Ruderman}}\ and\ \bibinfo {author} {\bibfnamefont {C.}~\bibnamefont
  {Kittel}},\ }\href {https://doi.org/10.1103/PhysRev.96.99} {\bibfield
  {journal} {\bibinfo  {journal} {Phys. Rev.}\ }\textbf {\bibinfo {volume}
  {96}},\ \bibinfo {pages} {99} (\bibinfo {year} {1954})}\BibitemShut {NoStop}%
\bibitem [{\citenamefont {{Kasuya}}(1956)}]{1956PThPh1645K}%
  \BibitemOpen
  \bibfield  {author} {\bibinfo {author} {\bibfnamefont {T.}~\bibnamefont
  {{Kasuya}}},\ }\href {https://doi.org/10.1143/PTP.16.45} {\bibfield
  {journal} {\bibinfo  {journal} {Progress of Theoretical Physics}\ }\textbf
  {\bibinfo {volume} {16}},\ \bibinfo {pages} {45} (\bibinfo {year}
  {1956})}\BibitemShut {NoStop}%
\bibitem [{\citenamefont {Yosida}(1957)}]{PhysRev.106.893}%
  \BibitemOpen
  \bibfield  {author} {\bibinfo {author} {\bibfnamefont {K.}~\bibnamefont
  {Yosida}},\ }\href {https://doi.org/10.1103/PhysRev.106.893} {\bibfield
  {journal} {\bibinfo  {journal} {Phys. Rev.}\ }\textbf {\bibinfo {volume}
  {106}},\ \bibinfo {pages} {893} (\bibinfo {year} {1957})}\BibitemShut
  {NoStop}%
\bibitem [{\citenamefont {Lavagna}\ and\ \citenamefont
  {Pepin}(1998)}]{Pepin_Kondo}%
  \BibitemOpen
  \bibfield  {author} {\bibinfo {author} {\bibfnamefont {M.}~\bibnamefont
  {Lavagna}}\ and\ \bibinfo {author} {\bibfnamefont {C.}~\bibnamefont
  {Pepin}},\ }\href@noop {} {\bibfield  {journal} {\bibinfo  {journal} {Acta
  Phys. Pol. B}\ }\textbf {\bibinfo {volume} {29}},\ \bibinfo {pages} {3753}
  (\bibinfo {year} {1998})}\BibitemShut {NoStop}%
\bibitem [{\citenamefont {Gulacsi}(2006)}]{Gulacsi2006-GULTKL}%
  \BibitemOpen
  \bibfield  {author} {\bibinfo {author} {\bibfnamefont {M.}~\bibnamefont
  {Gulacsi}},\ }\href {https://doi.org/10.1080/14786430500355045} {\bibfield
  {journal} {\bibinfo  {journal} {Philosophical Magazine}\ }\textbf {\bibinfo
  {volume} {86}},\ \bibinfo {pages} {1907} (\bibinfo {year}
  {2006})}\BibitemShut {NoStop}%
\bibitem [{\citenamefont {Doniach}(1977{\natexlab{a}})}]{DONIACH1977231}%
  \BibitemOpen
  \bibfield  {author} {\bibinfo {author} {\bibfnamefont {S.}~\bibnamefont
  {Doniach}},\ }\href
  {https://doi.org/https://doi.org/10.1016/0378-4363(77)90190-5} {\bibfield
  {journal} {\bibinfo  {journal} {Physica B+C}\ }\textbf {\bibinfo {volume}
  {91}},\ \bibinfo {pages} {231} (\bibinfo {year}
  {1977}{\natexlab{a}})}\BibitemShut {NoStop}%
\bibitem [{\citenamefont {Doniach}(1977{\natexlab{b}})}]{Doniach1977}%
  \BibitemOpen
  \bibfield  {author} {\bibinfo {author} {\bibfnamefont {S.}~\bibnamefont
  {Doniach}},\ }\bibinfo {title} {Phase diagram for the kondo lattice},\ in\
  \href {https://doi.org/10.1007/978-1-4615-8816-0_15} {\emph {\bibinfo
  {booktitle} {Valence Instabilities and Related Narrow-Band Phenomena}}},\
  \bibinfo {editor} {edited by\ \bibinfo {editor} {\bibfnamefont {R.~D.}\
  \bibnamefont {Parks}}}\ (\bibinfo  {publisher} {Springer US},\ \bibinfo
  {address} {Boston, MA},\ \bibinfo {year} {1977})\ pp.\ \bibinfo {pages}
  {169--176}\BibitemShut {NoStop}%
\bibitem [{\citenamefont {Rubtsov}\ \emph {et~al.}(2009)\citenamefont
  {Rubtsov}, \citenamefont {Katsnelson}, \citenamefont {Lichtenstein},\ and\
  \citenamefont {Georges}}]{PhysRevB.79.045133}%
  \BibitemOpen
  \bibfield  {author} {\bibinfo {author} {\bibfnamefont {A.~N.}\ \bibnamefont
  {Rubtsov}}, \bibinfo {author} {\bibfnamefont {M.~I.}\ \bibnamefont
  {Katsnelson}}, \bibinfo {author} {\bibfnamefont {A.~I.}\ \bibnamefont
  {Lichtenstein}},\ and\ \bibinfo {author} {\bibfnamefont {A.}~\bibnamefont
  {Georges}},\ }\href {https://doi.org/10.1103/PhysRevB.79.045133} {\bibfield
  {journal} {\bibinfo  {journal} {Phys. Rev. B}\ }\textbf {\bibinfo {volume}
  {79}},\ \bibinfo {pages} {045133} (\bibinfo {year} {2009})}\BibitemShut
  {NoStop}%
\bibitem [{\citenamefont {Hafermann}\ \emph {et~al.}(2009)\citenamefont
  {Hafermann}, \citenamefont {Li}, \citenamefont {Rubtsov}, \citenamefont
  {Katsnelson}, \citenamefont {Lichtenstein},\ and\ \citenamefont
  {Monien}}]{PhysRevLett.102.206401}%
  \BibitemOpen
  \bibfield  {author} {\bibinfo {author} {\bibfnamefont {H.}~\bibnamefont
  {Hafermann}}, \bibinfo {author} {\bibfnamefont {G.}~\bibnamefont {Li}},
  \bibinfo {author} {\bibfnamefont {A.~N.}\ \bibnamefont {Rubtsov}}, \bibinfo
  {author} {\bibfnamefont {M.~I.}\ \bibnamefont {Katsnelson}}, \bibinfo
  {author} {\bibfnamefont {A.~I.}\ \bibnamefont {Lichtenstein}},\ and\ \bibinfo
  {author} {\bibfnamefont {H.}~\bibnamefont {Monien}},\ }\href
  {https://doi.org/10.1103/PhysRevLett.102.206401} {\bibfield  {journal}
  {\bibinfo  {journal} {Phys. Rev. Lett.}\ }\textbf {\bibinfo {volume} {102}},\
  \bibinfo {pages} {206401} (\bibinfo {year} {2009})}\BibitemShut {NoStop}%
\bibitem [{\citenamefont {Georges}\ \emph {et~al.}(1996)\citenamefont
  {Georges}, \citenamefont {Kotliar}, \citenamefont {Krauth},\ and\
  \citenamefont {Rozenberg}}]{georges1996dynamical}%
  \BibitemOpen
  \bibfield  {author} {\bibinfo {author} {\bibfnamefont {A.}~\bibnamefont
  {Georges}}, \bibinfo {author} {\bibfnamefont {G.}~\bibnamefont {Kotliar}},
  \bibinfo {author} {\bibfnamefont {W.}~\bibnamefont {Krauth}},\ and\ \bibinfo
  {author} {\bibfnamefont {M.~J.}\ \bibnamefont {Rozenberg}},\ }\href
  {https://doi.org/10.1103/RevModPhys.68.13} {\bibfield  {journal} {\bibinfo
  {journal} {Rev. Mod. Phys.}\ }\textbf {\bibinfo {volume} {68}},\ \bibinfo
  {pages} {13} (\bibinfo {year} {1996})}\BibitemShut {NoStop}%
\bibitem [{\citenamefont {Burdin}\ \emph {et~al.}(2002)\citenamefont {Burdin},
  \citenamefont {Grempel},\ and\ \citenamefont {Georges}}]{burdin2002heavy}%
  \BibitemOpen
  \bibfield  {author} {\bibinfo {author} {\bibfnamefont {S.}~\bibnamefont
  {Burdin}}, \bibinfo {author} {\bibfnamefont {D.~R.}\ \bibnamefont
  {Grempel}},\ and\ \bibinfo {author} {\bibfnamefont {A.}~\bibnamefont
  {Georges}},\ }\href {https://doi.org/10.1103/PhysRevB.66.045111} {\bibfield
  {journal} {\bibinfo  {journal} {Phys. Rev. B}\ }\textbf {\bibinfo {volume}
  {66}},\ \bibinfo {pages} {045111} (\bibinfo {year} {2002})}\BibitemShut
  {NoStop}%
\bibitem [{\citenamefont {Otsuki}\ \emph
  {et~al.}(2009{\natexlab{a}})\citenamefont {Otsuki}, \citenamefont
  {Kusunose},\ and\ \citenamefont {Kuramoto}}]{otsuki2009evolution}%
  \BibitemOpen
  \bibfield  {author} {\bibinfo {author} {\bibfnamefont {J.}~\bibnamefont
  {Otsuki}}, \bibinfo {author} {\bibfnamefont {H.}~\bibnamefont {Kusunose}},\
  and\ \bibinfo {author} {\bibfnamefont {Y.}~\bibnamefont {Kuramoto}},\ }\href
  {https://doi.org/10.1103/PhysRevLett.102.017202} {\bibfield  {journal}
  {\bibinfo  {journal} {Phys. Rev. Lett.}\ }\textbf {\bibinfo {volume} {102}},\
  \bibinfo {pages} {017202} (\bibinfo {year} {2009}{\natexlab{a}})}\BibitemShut
  {NoStop}%
\bibitem [{\citenamefont {Peters}\ \emph {et~al.}(2013)\citenamefont {Peters},
  \citenamefont {Hoshino}, \citenamefont {Kawakami}, \citenamefont {Otsuki},\
  and\ \citenamefont {Kuramoto}}]{peters2013charge}%
  \BibitemOpen
  \bibfield  {author} {\bibinfo {author} {\bibfnamefont {R.}~\bibnamefont
  {Peters}}, \bibinfo {author} {\bibfnamefont {S.}~\bibnamefont {Hoshino}},
  \bibinfo {author} {\bibfnamefont {N.}~\bibnamefont {Kawakami}}, \bibinfo
  {author} {\bibfnamefont {J.}~\bibnamefont {Otsuki}},\ and\ \bibinfo {author}
  {\bibfnamefont {Y.}~\bibnamefont {Kuramoto}},\ }\href
  {https://doi.org/10.1103/PhysRevB.87.165133} {\bibfield  {journal} {\bibinfo
  {journal} {Phys. Rev. B}\ }\textbf {\bibinfo {volume} {87}},\ \bibinfo
  {pages} {165133} (\bibinfo {year} {2013})}\BibitemShut {NoStop}%
\bibitem [{\citenamefont {Sch\"afer}\ \emph {et~al.}(2021)\citenamefont
  {Sch\"afer}, \citenamefont {Wentzell}, \citenamefont {\ifmmode~\check{S}\else
  \v{S}\fi{}imkovic}, \citenamefont {He}, \citenamefont {Hille}, \citenamefont
  {Klett}, \citenamefont {Eckhardt}, \citenamefont {Arzhang}, \citenamefont
  {Harkov}, \citenamefont {Le~R\'egent}, \citenamefont {Kirsch}, \citenamefont
  {Wang}, \citenamefont {Kim}, \citenamefont {Kozik}, \citenamefont {Stepanov},
  \citenamefont {Kauch}, \citenamefont {Andergassen}, \citenamefont {Hansmann},
  \citenamefont {Rohe}, \citenamefont {Vilk}, \citenamefont {LeBlanc},
  \citenamefont {Zhang}, \citenamefont {Tremblay}, \citenamefont {Ferrero},
  \citenamefont {Parcollet},\ and\ \citenamefont
  {Georges}}]{PhysRevX.11.011058}%
  \BibitemOpen
  \bibfield  {author} {\bibinfo {author} {\bibfnamefont {T.}~\bibnamefont
  {Sch\"afer}}, \bibinfo {author} {\bibfnamefont {N.}~\bibnamefont {Wentzell}},
  \bibinfo {author} {\bibfnamefont {F.}~\bibnamefont {\ifmmode~\check{S}\else
  \v{S}\fi{}imkovic}}, \bibinfo {author} {\bibfnamefont {Y.-Y.}\ \bibnamefont
  {He}}, \bibinfo {author} {\bibfnamefont {C.}~\bibnamefont {Hille}}, \bibinfo
  {author} {\bibfnamefont {M.}~\bibnamefont {Klett}}, \bibinfo {author}
  {\bibfnamefont {C.~J.}\ \bibnamefont {Eckhardt}}, \bibinfo {author}
  {\bibfnamefont {B.}~\bibnamefont {Arzhang}}, \bibinfo {author} {\bibfnamefont
  {V.}~\bibnamefont {Harkov}}, \bibinfo {author} {\bibfnamefont {F.~m. c.-M.}\
  \bibnamefont {Le~R\'egent}}, \bibinfo {author} {\bibfnamefont
  {A.}~\bibnamefont {Kirsch}}, \bibinfo {author} {\bibfnamefont
  {Y.}~\bibnamefont {Wang}}, \bibinfo {author} {\bibfnamefont {A.~J.}\
  \bibnamefont {Kim}}, \bibinfo {author} {\bibfnamefont {E.}~\bibnamefont
  {Kozik}}, \bibinfo {author} {\bibfnamefont {E.~A.}\ \bibnamefont {Stepanov}},
  \bibinfo {author} {\bibfnamefont {A.}~\bibnamefont {Kauch}}, \bibinfo
  {author} {\bibfnamefont {S.}~\bibnamefont {Andergassen}}, \bibinfo {author}
  {\bibfnamefont {P.}~\bibnamefont {Hansmann}}, \bibinfo {author}
  {\bibfnamefont {D.}~\bibnamefont {Rohe}}, \bibinfo {author} {\bibfnamefont
  {Y.~M.}\ \bibnamefont {Vilk}}, \bibinfo {author} {\bibfnamefont {J.~P.~F.}\
  \bibnamefont {LeBlanc}}, \bibinfo {author} {\bibfnamefont {S.}~\bibnamefont
  {Zhang}}, \bibinfo {author} {\bibfnamefont {A.-M.~S.}\ \bibnamefont
  {Tremblay}}, \bibinfo {author} {\bibfnamefont {M.}~\bibnamefont {Ferrero}},
  \bibinfo {author} {\bibfnamefont {O.}~\bibnamefont {Parcollet}},\ and\
  \bibinfo {author} {\bibfnamefont {A.}~\bibnamefont {Georges}},\ }\href
  {https://doi.org/10.1103/PhysRevX.11.011058} {\bibfield  {journal} {\bibinfo
  {journal} {Phys. Rev. X}\ }\textbf {\bibinfo {volume} {11}},\ \bibinfo
  {pages} {011058} (\bibinfo {year} {2021})}\BibitemShut {NoStop}%
\bibitem [{\citenamefont {Fazekas}\ and\ \citenamefont
  {M{\"u}ller-Hartmann}(1991)}]{fazekas1991magnetic}%
  \BibitemOpen
  \bibfield  {author} {\bibinfo {author} {\bibfnamefont {P.}~\bibnamefont
  {Fazekas}}\ and\ \bibinfo {author} {\bibfnamefont {E.}~\bibnamefont
  {M{\"u}ller-Hartmann}},\ }\href {https://doi.org/10.1007/BF01313231}
  {\bibfield  {journal} {\bibinfo  {journal} {Zeitschrift f{\"u}r Physik B
  Condensed Matter}\ }\textbf {\bibinfo {volume} {85}},\ \bibinfo {pages} {285}
  (\bibinfo {year} {1991})}\BibitemShut {NoStop}%
\bibitem [{\citenamefont {Lacroix}\ and\ \citenamefont
  {Cyrot}(1979)}]{lacroix1979phase}%
  \BibitemOpen
  \bibfield  {author} {\bibinfo {author} {\bibfnamefont {C.}~\bibnamefont
  {Lacroix}}\ and\ \bibinfo {author} {\bibfnamefont {M.}~\bibnamefont
  {Cyrot}},\ }\href {https://doi.org/10.1103/PhysRevB.20.1969} {\bibfield
  {journal} {\bibinfo  {journal} {Phys. Rev. B}\ }\textbf {\bibinfo {volume}
  {20}},\ \bibinfo {pages} {1969} (\bibinfo {year} {1979})}\BibitemShut
  {NoStop}%
\bibitem [{\citenamefont {Martin}\ \emph {et~al.}(2010)\citenamefont {Martin},
  \citenamefont {Bercx},\ and\ \citenamefont {Assaad}}]{martin2010fermi}%
  \BibitemOpen
  \bibfield  {author} {\bibinfo {author} {\bibfnamefont {L.~C.}\ \bibnamefont
  {Martin}}, \bibinfo {author} {\bibfnamefont {M.}~\bibnamefont {Bercx}},\ and\
  \bibinfo {author} {\bibfnamefont {F.~F.}\ \bibnamefont {Assaad}},\ }\href
  {https://doi.org/10.1103/PhysRevB.82.245105} {\bibfield  {journal} {\bibinfo
  {journal} {Phys. Rev. B}\ }\textbf {\bibinfo {volume} {82}},\ \bibinfo
  {pages} {245105} (\bibinfo {year} {2010})}\BibitemShut {NoStop}%
\bibitem [{\citenamefont {Martin}\ and\ \citenamefont
  {Assaad}(2008{\natexlab{a}})}]{martin2008evolution}%
  \BibitemOpen
  \bibfield  {author} {\bibinfo {author} {\bibfnamefont {L.~C.}\ \bibnamefont
  {Martin}}\ and\ \bibinfo {author} {\bibfnamefont {F.~F.}\ \bibnamefont
  {Assaad}},\ }\href {https://doi.org/10.1103/PhysRevLett.101.066404}
  {\bibfield  {journal} {\bibinfo  {journal} {Phys. Rev. Lett.}\ }\textbf
  {\bibinfo {volume} {101}},\ \bibinfo {pages} {066404} (\bibinfo {year}
  {2008}{\natexlab{a}})}\BibitemShut {NoStop}%
\bibitem [{\citenamefont {Otsuki}\ \emph
  {et~al.}(2009{\natexlab{b}})\citenamefont {Otsuki}, \citenamefont
  {Kusunose},\ and\ \citenamefont {Kuramoto}}]{doi:10.1143/JPSJ.78.014702}%
  \BibitemOpen
  \bibfield  {author} {\bibinfo {author} {\bibfnamefont {J.}~\bibnamefont
  {Otsuki}}, \bibinfo {author} {\bibfnamefont {H.}~\bibnamefont {Kusunose}},\
  and\ \bibinfo {author} {\bibfnamefont {Y.}~\bibnamefont {Kuramoto}},\ }\href
  {https://doi.org/10.1143/JPSJ.78.014702} {\bibfield  {journal} {\bibinfo
  {journal} {Journal of the Physical Society of Japan}\ }\textbf {\bibinfo
  {volume} {78}},\ \bibinfo {pages} {014702} (\bibinfo {year}
  {2009}{\natexlab{b}})}\BibitemShut {NoStop}%
\bibitem [{\citenamefont {Otsuki}\ \emph
  {et~al.}(2009{\natexlab{c}})\citenamefont {Otsuki}, \citenamefont
  {Kusunose},\ and\ \citenamefont {Kuramoto}}]{doi:10.1143/JPSJ.78.034719}%
  \BibitemOpen
  \bibfield  {author} {\bibinfo {author} {\bibfnamefont {J.}~\bibnamefont
  {Otsuki}}, \bibinfo {author} {\bibfnamefont {H.}~\bibnamefont {Kusunose}},\
  and\ \bibinfo {author} {\bibfnamefont {Y.}~\bibnamefont {Kuramoto}},\ }\href
  {https://doi.org/10.1143/JPSJ.78.034719} {\bibfield  {journal} {\bibinfo
  {journal} {Journal of the Physical Society of Japan}\ }\textbf {\bibinfo
  {volume} {78}},\ \bibinfo {pages} {034719} (\bibinfo {year}
  {2009}{\natexlab{c}})}\BibitemShut {NoStop}%
\bibitem [{\citenamefont {Akagi}\ \emph {et~al.}(2012)\citenamefont {Akagi},
  \citenamefont {Udagawa},\ and\ \citenamefont {Motome}}]{akagi2012hidden}%
  \BibitemOpen
  \bibfield  {author} {\bibinfo {author} {\bibfnamefont {Y.}~\bibnamefont
  {Akagi}}, \bibinfo {author} {\bibfnamefont {M.}~\bibnamefont {Udagawa}},\
  and\ \bibinfo {author} {\bibfnamefont {Y.}~\bibnamefont {Motome}},\ }\href
  {https://doi.org/10.1103/PhysRevLett.108.096401} {\bibfield  {journal}
  {\bibinfo  {journal} {Phys. Rev. Lett.}\ }\textbf {\bibinfo {volume} {108}},\
  \bibinfo {pages} {096401} (\bibinfo {year} {2012})}\BibitemShut {NoStop}%
\bibitem [{\citenamefont {Aulbach}\ \emph {et~al.}(2015)\citenamefont
  {Aulbach}, \citenamefont {Assaad},\ and\ \citenamefont
  {Potthoff}}]{aulbach2015dynamical}%
  \BibitemOpen
  \bibfield  {author} {\bibinfo {author} {\bibfnamefont {M.~W.}\ \bibnamefont
  {Aulbach}}, \bibinfo {author} {\bibfnamefont {F.~F.}\ \bibnamefont
  {Assaad}},\ and\ \bibinfo {author} {\bibfnamefont {M.}~\bibnamefont
  {Potthoff}},\ }\href {https://doi.org/10.1103/PhysRevB.92.235131} {\bibfield
  {journal} {\bibinfo  {journal} {Phys. Rev. B}\ }\textbf {\bibinfo {volume}
  {92}},\ \bibinfo {pages} {235131} (\bibinfo {year} {2015})}\BibitemShut
  {NoStop}%
\bibitem [{\citenamefont {Ke\ss{}ler}\ and\ \citenamefont
  {Eder}(2020)}]{kessler2020magnetic}%
  \BibitemOpen
  \bibfield  {author} {\bibinfo {author} {\bibfnamefont {M.}~\bibnamefont
  {Ke\ss{}ler}}\ and\ \bibinfo {author} {\bibfnamefont {R.}~\bibnamefont
  {Eder}},\ }\href {https://doi.org/10.1103/PhysRevB.102.235125} {\bibfield
  {journal} {\bibinfo  {journal} {Phys. Rev. B}\ }\textbf {\bibinfo {volume}
  {102}},\ \bibinfo {pages} {235125} (\bibinfo {year} {2020})}\BibitemShut
  {NoStop}%
\bibitem [{\citenamefont {Inui}\ and\ \citenamefont
  {Motome}(2020)}]{PhysRevB.102.155126}%
  \BibitemOpen
  \bibfield  {author} {\bibinfo {author} {\bibfnamefont {K.}~\bibnamefont
  {Inui}}\ and\ \bibinfo {author} {\bibfnamefont {Y.}~\bibnamefont {Motome}},\
  }\href {https://doi.org/10.1103/PhysRevB.102.155126} {\bibfield  {journal}
  {\bibinfo  {journal} {Phys. Rev. B}\ }\textbf {\bibinfo {volume} {102}},\
  \bibinfo {pages} {155126} (\bibinfo {year} {2020})}\BibitemShut {NoStop}%
\bibitem [{\citenamefont {Huang}\ \emph {et~al.}(2017)\citenamefont {Huang},
  \citenamefont {Clark}, \citenamefont {Navarro-Moratalla}, \citenamefont
  {Klein}, \citenamefont {Cheng}, \citenamefont {Seyler}, \citenamefont
  {Zhong}, \citenamefont {Schmidgall}, \citenamefont {McGuire}, \citenamefont
  {Cobden}, \citenamefont {Yao}, \citenamefont {Xiao}, \citenamefont
  {Jarillo-Herrero},\ and\ \citenamefont {Xu}}]{Nature_CrI3}%
  \BibitemOpen
  \bibfield  {author} {\bibinfo {author} {\bibfnamefont {B.}~\bibnamefont
  {Huang}}, \bibinfo {author} {\bibfnamefont {G.}~\bibnamefont {Clark}},
  \bibinfo {author} {\bibfnamefont {E.}~\bibnamefont {Navarro-Moratalla}},
  \bibinfo {author} {\bibfnamefont {D.~R.}\ \bibnamefont {Klein}}, \bibinfo
  {author} {\bibfnamefont {R.}~\bibnamefont {Cheng}}, \bibinfo {author}
  {\bibfnamefont {K.~L.}\ \bibnamefont {Seyler}}, \bibinfo {author}
  {\bibfnamefont {D.}~\bibnamefont {Zhong}}, \bibinfo {author} {\bibfnamefont
  {E.}~\bibnamefont {Schmidgall}}, \bibinfo {author} {\bibfnamefont {M.~A.}\
  \bibnamefont {McGuire}}, \bibinfo {author} {\bibfnamefont {D.~H.}\
  \bibnamefont {Cobden}}, \bibinfo {author} {\bibfnamefont {W.}~\bibnamefont
  {Yao}}, \bibinfo {author} {\bibfnamefont {D.}~\bibnamefont {Xiao}}, \bibinfo
  {author} {\bibfnamefont {P.}~\bibnamefont {Jarillo-Herrero}},\ and\ \bibinfo
  {author} {\bibfnamefont {X.}~\bibnamefont {Xu}},\ }\href@noop {} {\bibfield
  {journal} {\bibinfo  {journal} {Nature}\ }\textbf {\bibinfo {volume} {546}},\
  \bibinfo {pages} {270} (\bibinfo {year} {2017})}\BibitemShut {NoStop}%
\bibitem [{\citenamefont {Gong}\ \emph {et~al.}(2017)\citenamefont {Gong},
  \citenamefont {Li}, \citenamefont {Li}, \citenamefont {Ji}, \citenamefont
  {Stern}, \citenamefont {Xia}, \citenamefont {Cao}, \citenamefont {Bao},
  \citenamefont {Wang}, \citenamefont {Wang}, \citenamefont {Qiu},
  \citenamefont {Cava}, \citenamefont {Louie}, \citenamefont {Xia},\ and\
  \citenamefont {Zhang}}]{Gong_Discovery_2017}%
  \BibitemOpen
  \bibfield  {author} {\bibinfo {author} {\bibfnamefont {C.}~\bibnamefont
  {Gong}}, \bibinfo {author} {\bibfnamefont {L.}~\bibnamefont {Li}}, \bibinfo
  {author} {\bibfnamefont {Z.}~\bibnamefont {Li}}, \bibinfo {author}
  {\bibfnamefont {H.}~\bibnamefont {Ji}}, \bibinfo {author} {\bibfnamefont
  {A.}~\bibnamefont {Stern}}, \bibinfo {author} {\bibfnamefont
  {Y.}~\bibnamefont {Xia}}, \bibinfo {author} {\bibfnamefont {T.}~\bibnamefont
  {Cao}}, \bibinfo {author} {\bibfnamefont {W.}~\bibnamefont {Bao}}, \bibinfo
  {author} {\bibfnamefont {C.}~\bibnamefont {Wang}}, \bibinfo {author}
  {\bibfnamefont {Y.}~\bibnamefont {Wang}}, \bibinfo {author} {\bibfnamefont
  {Z.}~\bibnamefont {Qiu}}, \bibinfo {author} {\bibfnamefont {R.}~\bibnamefont
  {Cava}}, \bibinfo {author} {\bibfnamefont {S.~G.}\ \bibnamefont {Louie}},
  \bibinfo {author} {\bibfnamefont {J.}~\bibnamefont {Xia}},\ and\ \bibinfo
  {author} {\bibfnamefont {X.}~\bibnamefont {Zhang}},\ }\href
  {https://doi.org/10.1038/nature22060} {\bibfield  {journal} {\bibinfo
  {journal} {Nature}\ }\textbf {\bibinfo {volume} {546}},\ \bibinfo {pages}
  {265} (\bibinfo {year} {2017})}\BibitemShut {NoStop}%
\bibitem [{\citenamefont {Zhang}\ \emph {et~al.}(2019)\citenamefont {Zhang},
  \citenamefont {Cai}, \citenamefont {Xia}, \citenamefont {Liang},
  \citenamefont {Huang}, \citenamefont {Wang}, \citenamefont {Yang},
  \citenamefont {Yuan}, \citenamefont {Chen}, \citenamefont {Zhang},
  \citenamefont {Guo}, \citenamefont {Liu},\ and\ \citenamefont
  {Li}}]{PhysRevLett.123.047203}%
  \BibitemOpen
  \bibfield  {author} {\bibinfo {author} {\bibfnamefont {J.}~\bibnamefont
  {Zhang}}, \bibinfo {author} {\bibfnamefont {X.}~\bibnamefont {Cai}}, \bibinfo
  {author} {\bibfnamefont {W.}~\bibnamefont {Xia}}, \bibinfo {author}
  {\bibfnamefont {A.}~\bibnamefont {Liang}}, \bibinfo {author} {\bibfnamefont
  {J.}~\bibnamefont {Huang}}, \bibinfo {author} {\bibfnamefont
  {C.}~\bibnamefont {Wang}}, \bibinfo {author} {\bibfnamefont {L.}~\bibnamefont
  {Yang}}, \bibinfo {author} {\bibfnamefont {H.}~\bibnamefont {Yuan}}, \bibinfo
  {author} {\bibfnamefont {Y.}~\bibnamefont {Chen}}, \bibinfo {author}
  {\bibfnamefont {S.}~\bibnamefont {Zhang}}, \bibinfo {author} {\bibfnamefont
  {Y.}~\bibnamefont {Guo}}, \bibinfo {author} {\bibfnamefont {Z.}~\bibnamefont
  {Liu}},\ and\ \bibinfo {author} {\bibfnamefont {G.}~\bibnamefont {Li}},\
  }\href {https://doi.org/10.1103/PhysRevLett.123.047203} {\bibfield  {journal}
  {\bibinfo  {journal} {Phys. Rev. Lett.}\ }\textbf {\bibinfo {volume} {123}},\
  \bibinfo {pages} {047203} (\bibinfo {year} {2019})}\BibitemShut {NoStop}%
\bibitem [{\citenamefont {Mermin}\ and\ \citenamefont
  {Wagner}(1966)}]{PhysRevLett.17.1133}%
  \BibitemOpen
  \bibfield  {author} {\bibinfo {author} {\bibfnamefont {N.~D.}\ \bibnamefont
  {Mermin}}\ and\ \bibinfo {author} {\bibfnamefont {H.}~\bibnamefont
  {Wagner}},\ }\href {https://doi.org/10.1103/PhysRevLett.17.1133} {\bibfield
  {journal} {\bibinfo  {journal} {Phys. Rev. Lett.}\ }\textbf {\bibinfo
  {volume} {17}},\ \bibinfo {pages} {1133} (\bibinfo {year}
  {1966})}\BibitemShut {NoStop}%
\bibitem [{\citenamefont {Coleman}(1983)}]{PhysRevB.28.5255}%
  \BibitemOpen
  \bibfield  {author} {\bibinfo {author} {\bibfnamefont {P.}~\bibnamefont
  {Coleman}},\ }\href {https://doi.org/10.1103/PhysRevB.28.5255} {\bibfield
  {journal} {\bibinfo  {journal} {Phys. Rev. B}\ }\textbf {\bibinfo {volume}
  {28}},\ \bibinfo {pages} {5255} (\bibinfo {year} {1983})}\BibitemShut
  {NoStop}%
\bibitem [{\citenamefont {Hoshino}\ \emph {et~al.}(2010)\citenamefont
  {Hoshino}, \citenamefont {Otsuki},\ and\ \citenamefont
  {Kuramoto}}]{PhysRevB.81.113108}%
  \BibitemOpen
  \bibfield  {author} {\bibinfo {author} {\bibfnamefont {S.}~\bibnamefont
  {Hoshino}}, \bibinfo {author} {\bibfnamefont {J.}~\bibnamefont {Otsuki}},\
  and\ \bibinfo {author} {\bibfnamefont {Y.}~\bibnamefont {Kuramoto}},\ }\href
  {https://doi.org/10.1103/PhysRevB.81.113108} {\bibfield  {journal} {\bibinfo
  {journal} {Phys. Rev. B}\ }\textbf {\bibinfo {volume} {81}},\ \bibinfo
  {pages} {113108} (\bibinfo {year} {2010})}\BibitemShut {NoStop}%
\bibitem [{\citenamefont {Watanabe}\ and\ \citenamefont
  {Ogata}(2007)}]{PhysRevLett.99.136401}%
  \BibitemOpen
  \bibfield  {author} {\bibinfo {author} {\bibfnamefont {H.}~\bibnamefont
  {Watanabe}}\ and\ \bibinfo {author} {\bibfnamefont {M.}~\bibnamefont
  {Ogata}},\ }\href {https://doi.org/10.1103/PhysRevLett.99.136401} {\bibfield
  {journal} {\bibinfo  {journal} {Phys. Rev. Lett.}\ }\textbf {\bibinfo
  {volume} {99}},\ \bibinfo {pages} {136401} (\bibinfo {year}
  {2007})}\BibitemShut {NoStop}%
\bibitem [{\citenamefont {Martin}\ and\ \citenamefont
  {Assaad}(2008{\natexlab{b}})}]{PhysRevLett.101.066404}%
  \BibitemOpen
  \bibfield  {author} {\bibinfo {author} {\bibfnamefont {L.~C.}\ \bibnamefont
  {Martin}}\ and\ \bibinfo {author} {\bibfnamefont {F.~F.}\ \bibnamefont
  {Assaad}},\ }\href {https://doi.org/10.1103/PhysRevLett.101.066404}
  {\bibfield  {journal} {\bibinfo  {journal} {Phys. Rev. Lett.}\ }\textbf
  {\bibinfo {volume} {101}},\ \bibinfo {pages} {066404} (\bibinfo {year}
  {2008}{\natexlab{b}})}\BibitemShut {NoStop}%
\bibitem [{\citenamefont {Danu}\ \emph {et~al.}(2021)\citenamefont {Danu},
  \citenamefont {Liu}, \citenamefont {Assaad},\ and\ \citenamefont
  {Raczkowski}}]{PhysRevB.104.155128}%
  \BibitemOpen
  \bibfield  {author} {\bibinfo {author} {\bibfnamefont {B.}~\bibnamefont
  {Danu}}, \bibinfo {author} {\bibfnamefont {Z.}~\bibnamefont {Liu}}, \bibinfo
  {author} {\bibfnamefont {F.~F.}\ \bibnamefont {Assaad}},\ and\ \bibinfo
  {author} {\bibfnamefont {M.}~\bibnamefont {Raczkowski}},\ }\href
  {https://doi.org/10.1103/PhysRevB.104.155128} {\bibfield  {journal} {\bibinfo
   {journal} {Phys. Rev. B}\ }\textbf {\bibinfo {volume} {104}},\ \bibinfo
  {pages} {155128} (\bibinfo {year} {2021})}\BibitemShut {NoStop}%
\bibitem [{\citenamefont {Otsuki}(2015)}]{PhysRevLett.115.036404}%
  \BibitemOpen
  \bibfield  {author} {\bibinfo {author} {\bibfnamefont {J.}~\bibnamefont
  {Otsuki}},\ }\href {https://doi.org/10.1103/PhysRevLett.115.036404}
  {\bibfield  {journal} {\bibinfo  {journal} {Phys. Rev. Lett.}\ }\textbf
  {\bibinfo {volume} {115}},\ \bibinfo {pages} {036404} (\bibinfo {year}
  {2015})}\BibitemShut {NoStop}%
\bibitem [{\citenamefont {Kyung}\ and\ \citenamefont
  {Tremblay}(2006)}]{PhysRevLett.97.046402}%
  \BibitemOpen
  \bibfield  {author} {\bibinfo {author} {\bibfnamefont {B.}~\bibnamefont
  {Kyung}}\ and\ \bibinfo {author} {\bibfnamefont {A.-M.~S.}\ \bibnamefont
  {Tremblay}},\ }\href {https://doi.org/10.1103/PhysRevLett.97.046402}
  {\bibfield  {journal} {\bibinfo  {journal} {Phys. Rev. Lett.}\ }\textbf
  {\bibinfo {volume} {97}},\ \bibinfo {pages} {046402} (\bibinfo {year}
  {2006})}\BibitemShut {NoStop}%
\bibitem [{\citenamefont {Li}\ \emph {et~al.}(2014)\citenamefont {Li},
  \citenamefont {Antipov}, \citenamefont {Rubtsov}, \citenamefont {Kirchner},\
  and\ \citenamefont {Hanke}}]{PhysRevB.89.161118}%
  \BibitemOpen
  \bibfield  {author} {\bibinfo {author} {\bibfnamefont {G.}~\bibnamefont
  {Li}}, \bibinfo {author} {\bibfnamefont {A.~E.}\ \bibnamefont {Antipov}},
  \bibinfo {author} {\bibfnamefont {A.~N.}\ \bibnamefont {Rubtsov}}, \bibinfo
  {author} {\bibfnamefont {S.}~\bibnamefont {Kirchner}},\ and\ \bibinfo
  {author} {\bibfnamefont {W.}~\bibnamefont {Hanke}},\ }\href
  {https://doi.org/10.1103/PhysRevB.89.161118} {\bibfield  {journal} {\bibinfo
  {journal} {Phys. Rev. B}\ }\textbf {\bibinfo {volume} {89}},\ \bibinfo
  {pages} {161118} (\bibinfo {year} {2014})}\BibitemShut {NoStop}%
\bibitem [{\citenamefont {Gao}\ \emph {et~al.}(2021)\citenamefont {Gao},
  \citenamefont {Hu}, \citenamefont {Sun}, \citenamefont {Wang}, \citenamefont
  {Lin},\ and\ \citenamefont {Li}}]{PhysRevB.103.235134}%
  \BibitemOpen
  \bibfield  {author} {\bibinfo {author} {\bibfnamefont {X.}~\bibnamefont
  {Gao}}, \bibinfo {author} {\bibfnamefont {C.}~\bibnamefont {Hu}}, \bibinfo
  {author} {\bibfnamefont {J.}~\bibnamefont {Sun}}, \bibinfo {author}
  {\bibfnamefont {X.-Q.}\ \bibnamefont {Wang}}, \bibinfo {author}
  {\bibfnamefont {H.-Q.}\ \bibnamefont {Lin}},\ and\ \bibinfo {author}
  {\bibfnamefont {G.}~\bibnamefont {Li}},\ }\href
  {https://doi.org/10.1103/PhysRevB.103.235134} {\bibfield  {journal} {\bibinfo
   {journal} {Phys. Rev. B}\ }\textbf {\bibinfo {volume} {103}},\ \bibinfo
  {pages} {235134} (\bibinfo {year} {2021})}\BibitemShut {NoStop}%
\bibitem [{\citenamefont {White}\ \emph {et~al.}(1989)\citenamefont {White},
  \citenamefont {Scalapino}, \citenamefont {Sugar}, \citenamefont {Loh},
  \citenamefont {Gubernatis},\ and\ \citenamefont
  {Scalettar}}]{PhysRevB.40.506}%
  \BibitemOpen
  \bibfield  {author} {\bibinfo {author} {\bibfnamefont {S.~R.}\ \bibnamefont
  {White}}, \bibinfo {author} {\bibfnamefont {D.~J.}\ \bibnamefont
  {Scalapino}}, \bibinfo {author} {\bibfnamefont {R.~L.}\ \bibnamefont
  {Sugar}}, \bibinfo {author} {\bibfnamefont {E.~Y.}\ \bibnamefont {Loh}},
  \bibinfo {author} {\bibfnamefont {J.~E.}\ \bibnamefont {Gubernatis}},\ and\
  \bibinfo {author} {\bibfnamefont {R.~T.}\ \bibnamefont {Scalettar}},\ }\href
  {https://doi.org/10.1103/PhysRevB.40.506} {\bibfield  {journal} {\bibinfo
  {journal} {Phys. Rev. B}\ }\textbf {\bibinfo {volume} {40}},\ \bibinfo
  {pages} {506} (\bibinfo {year} {1989})}\BibitemShut {NoStop}%
\bibitem [{\citenamefont {Gr\"ober}\ \emph {et~al.}(2000)\citenamefont
  {Gr\"ober}, \citenamefont {Eder},\ and\ \citenamefont
  {Hanke}}]{PhysRevB.62.4336}%
  \BibitemOpen
  \bibfield  {author} {\bibinfo {author} {\bibfnamefont {C.}~\bibnamefont
  {Gr\"ober}}, \bibinfo {author} {\bibfnamefont {R.}~\bibnamefont {Eder}},\
  and\ \bibinfo {author} {\bibfnamefont {W.}~\bibnamefont {Hanke}},\ }\href
  {https://doi.org/10.1103/PhysRevB.62.4336} {\bibfield  {journal} {\bibinfo
  {journal} {Phys. Rev. B}\ }\textbf {\bibinfo {volume} {62}},\ \bibinfo
  {pages} {4336} (\bibinfo {year} {2000})}\BibitemShut {NoStop}%
\bibitem [{\citenamefont {Lee}\ \emph {et~al.}(2008)\citenamefont {Lee},
  \citenamefont {Li},\ and\ \citenamefont {Monien}}]{PhysRevB.78.205117}%
  \BibitemOpen
  \bibfield  {author} {\bibinfo {author} {\bibfnamefont {H.}~\bibnamefont
  {Lee}}, \bibinfo {author} {\bibfnamefont {G.}~\bibnamefont {Li}},\ and\
  \bibinfo {author} {\bibfnamefont {H.}~\bibnamefont {Monien}},\ }\href
  {https://doi.org/10.1103/PhysRevB.78.205117} {\bibfield  {journal} {\bibinfo
  {journal} {Phys. Rev. B}\ }\textbf {\bibinfo {volume} {78}},\ \bibinfo
  {pages} {205117} (\bibinfo {year} {2008})}\BibitemShut {NoStop}%
\bibitem [{\citenamefont {Varney}\ \emph {et~al.}(2009)\citenamefont {Varney},
  \citenamefont {Lee}, \citenamefont {Bai}, \citenamefont {Chiesa},
  \citenamefont {Jarrell},\ and\ \citenamefont
  {Scalettar}}]{PhysRevB.80.075116}%
  \BibitemOpen
  \bibfield  {author} {\bibinfo {author} {\bibfnamefont {C.~N.}\ \bibnamefont
  {Varney}}, \bibinfo {author} {\bibfnamefont {C.-R.}\ \bibnamefont {Lee}},
  \bibinfo {author} {\bibfnamefont {Z.~J.}\ \bibnamefont {Bai}}, \bibinfo
  {author} {\bibfnamefont {S.}~\bibnamefont {Chiesa}}, \bibinfo {author}
  {\bibfnamefont {M.}~\bibnamefont {Jarrell}},\ and\ \bibinfo {author}
  {\bibfnamefont {R.~T.}\ \bibnamefont {Scalettar}},\ }\href
  {https://doi.org/10.1103/PhysRevB.80.075116} {\bibfield  {journal} {\bibinfo
  {journal} {Phys. Rev. B}\ }\textbf {\bibinfo {volume} {80}},\ \bibinfo
  {pages} {075116} (\bibinfo {year} {2009})}\BibitemShut {NoStop}%
\bibitem [{\citenamefont {Bulut}\ \emph {et~al.}(1994)\citenamefont {Bulut},
  \citenamefont {Scalapino},\ and\ \citenamefont {White}}]{PhysRevLett.72.705}%
  \BibitemOpen
  \bibfield  {author} {\bibinfo {author} {\bibfnamefont {N.}~\bibnamefont
  {Bulut}}, \bibinfo {author} {\bibfnamefont {D.~J.}\ \bibnamefont
  {Scalapino}},\ and\ \bibinfo {author} {\bibfnamefont {S.~R.}\ \bibnamefont
  {White}},\ }\href {https://doi.org/10.1103/PhysRevLett.72.705} {\bibfield
  {journal} {\bibinfo  {journal} {Phys. Rev. Lett.}\ }\textbf {\bibinfo
  {volume} {72}},\ \bibinfo {pages} {705} (\bibinfo {year} {1994})}\BibitemShut
  {NoStop}%
\bibitem [{\citenamefont {Laubach}\ \emph {et~al.}(2015)\citenamefont
  {Laubach}, \citenamefont {Thomale}, \citenamefont {Platt}, \citenamefont
  {Hanke},\ and\ \citenamefont {Li}}]{PhysRevB.91.245125}%
  \BibitemOpen
  \bibfield  {author} {\bibinfo {author} {\bibfnamefont {M.}~\bibnamefont
  {Laubach}}, \bibinfo {author} {\bibfnamefont {R.}~\bibnamefont {Thomale}},
  \bibinfo {author} {\bibfnamefont {C.}~\bibnamefont {Platt}}, \bibinfo
  {author} {\bibfnamefont {W.}~\bibnamefont {Hanke}},\ and\ \bibinfo {author}
  {\bibfnamefont {G.}~\bibnamefont {Li}},\ }\href
  {https://doi.org/10.1103/PhysRevB.91.245125} {\bibfield  {journal} {\bibinfo
  {journal} {Phys. Rev. B}\ }\textbf {\bibinfo {volume} {91}},\ \bibinfo
  {pages} {245125} (\bibinfo {year} {2015})}\BibitemShut {NoStop}%
\bibitem [{\citenamefont {Kokalj}\ and\ \citenamefont
  {McKenzie}(2015)}]{PhysRevB.91.205121}%
  \BibitemOpen
  \bibfield  {author} {\bibinfo {author} {\bibfnamefont {J.}~\bibnamefont
  {Kokalj}}\ and\ \bibinfo {author} {\bibfnamefont {R.~H.}\ \bibnamefont
  {McKenzie}},\ }\href {https://doi.org/10.1103/PhysRevB.91.205121} {\bibfield
  {journal} {\bibinfo  {journal} {Phys. Rev. B}\ }\textbf {\bibinfo {volume}
  {91}},\ \bibinfo {pages} {205121} (\bibinfo {year} {2015})}\BibitemShut
  {NoStop}%
\bibitem [{\citenamefont {Rohringer}\ and\ \citenamefont
  {Toschi}(2016)}]{PhysRevB.94.125144}%
  \BibitemOpen
  \bibfield  {author} {\bibinfo {author} {\bibfnamefont {G.}~\bibnamefont
  {Rohringer}}\ and\ \bibinfo {author} {\bibfnamefont {A.}~\bibnamefont
  {Toschi}},\ }\href {https://doi.org/10.1103/PhysRevB.94.125144} {\bibfield
  {journal} {\bibinfo  {journal} {Phys. Rev. B}\ }\textbf {\bibinfo {volume}
  {94}},\ \bibinfo {pages} {125144} (\bibinfo {year} {2016})}\BibitemShut
  {NoStop}%
\bibitem [{\citenamefont {Shirakawa}\ \emph {et~al.}(2017)\citenamefont
  {Shirakawa}, \citenamefont {Tohyama}, \citenamefont {Kokalj}, \citenamefont
  {Sota},\ and\ \citenamefont {Yunoki}}]{PhysRevB.96.205130}%
  \BibitemOpen
  \bibfield  {author} {\bibinfo {author} {\bibfnamefont {T.}~\bibnamefont
  {Shirakawa}}, \bibinfo {author} {\bibfnamefont {T.}~\bibnamefont {Tohyama}},
  \bibinfo {author} {\bibfnamefont {J.}~\bibnamefont {Kokalj}}, \bibinfo
  {author} {\bibfnamefont {S.}~\bibnamefont {Sota}},\ and\ \bibinfo {author}
  {\bibfnamefont {S.}~\bibnamefont {Yunoki}},\ }\href
  {https://doi.org/10.1103/PhysRevB.96.205130} {\bibfield  {journal} {\bibinfo
  {journal} {Phys. Rev. B}\ }\textbf {\bibinfo {volume} {96}},\ \bibinfo
  {pages} {205130} (\bibinfo {year} {2017})}\BibitemShut {NoStop}%
\bibitem [{\citenamefont {Qin}\ \emph {et~al.}(2022)\citenamefont {Qin},
  \citenamefont {Sch\"{a}fer}, \citenamefont {Andergassen}, \citenamefont
  {Corboz},\ and\ \citenamefont
  {Gull}}]{doi:10.1146/annurev-conmatphys-090921-033948}%
  \BibitemOpen
  \bibfield  {author} {\bibinfo {author} {\bibfnamefont {M.}~\bibnamefont
  {Qin}}, \bibinfo {author} {\bibfnamefont {T.}~\bibnamefont {Sch\"{a}fer}},
  \bibinfo {author} {\bibfnamefont {S.}~\bibnamefont {Andergassen}}, \bibinfo
  {author} {\bibfnamefont {P.}~\bibnamefont {Corboz}},\ and\ \bibinfo {author}
  {\bibfnamefont {E.}~\bibnamefont {Gull}},\ }\href
  {https://doi.org/10.1146/annurev-conmatphys-090921-033948} {\bibfield
  {journal} {\bibinfo  {journal} {Annual Review of Condensed Matter Physics}\
  }\textbf {\bibinfo {volume} {13}},\ \bibinfo {pages} {275} (\bibinfo {year}
  {2022})}\BibitemShut {NoStop}%
\bibitem [{\citenamefont {Sahebsara}\ and\ \citenamefont
  {S\'en\'echal}(2008)}]{PhysRevLett.100.136402}%
  \BibitemOpen
  \bibfield  {author} {\bibinfo {author} {\bibfnamefont {P.}~\bibnamefont
  {Sahebsara}}\ and\ \bibinfo {author} {\bibfnamefont {D.}~\bibnamefont
  {S\'en\'echal}},\ }\href {https://doi.org/10.1103/PhysRevLett.100.136402}
  {\bibfield  {journal} {\bibinfo  {journal} {Phys. Rev. Lett.}\ }\textbf
  {\bibinfo {volume} {100}},\ \bibinfo {pages} {136402} (\bibinfo {year}
  {2008})}\BibitemShut {NoStop}%
\bibitem [{\citenamefont {Li}\ and\ \citenamefont
  {Gull}(2020)}]{PhysRevResearch.2.013295}%
  \BibitemOpen
  \bibfield  {author} {\bibinfo {author} {\bibfnamefont {S.}~\bibnamefont
  {Li}}\ and\ \bibinfo {author} {\bibfnamefont {E.}~\bibnamefont {Gull}},\
  }\href {https://doi.org/10.1103/PhysRevResearch.2.013295} {\bibfield
  {journal} {\bibinfo  {journal} {Phys. Rev. Res.}\ }\textbf {\bibinfo {volume}
  {2}},\ \bibinfo {pages} {013295} (\bibinfo {year} {2020})}\BibitemShut
  {NoStop}%
\bibitem [{\citenamefont {Maier}\ \emph {et~al.}(2005)\citenamefont {Maier},
  \citenamefont {Jarrell}, \citenamefont {Pruschke},\ and\ \citenamefont
  {Hettler}}]{maier2005quantum}%
  \BibitemOpen
  \bibfield  {author} {\bibinfo {author} {\bibfnamefont {T.}~\bibnamefont
  {Maier}}, \bibinfo {author} {\bibfnamefont {M.}~\bibnamefont {Jarrell}},
  \bibinfo {author} {\bibfnamefont {T.}~\bibnamefont {Pruschke}},\ and\
  \bibinfo {author} {\bibfnamefont {M.~H.}\ \bibnamefont {Hettler}},\ }\href
  {https://doi.org/10.1103/RevModPhys.77.1027} {\bibfield  {journal} {\bibinfo
  {journal} {Rev. Mod. Phys.}\ }\textbf {\bibinfo {volume} {77}},\ \bibinfo
  {pages} {1027} (\bibinfo {year} {2005})}\BibitemShut {NoStop}%
\bibitem [{\citenamefont {Li}(2013)}]{li2013kondo}%
  \BibitemOpen
  \bibfield  {author} {\bibinfo {author} {\bibfnamefont {G.}~\bibnamefont
  {Li}},\ }\href {https://arxiv.org/abs/1309.0156} {\bibfield  {journal}
  {\bibinfo  {journal} {arXiv preprint arXiv:1309.0156}\ } (\bibinfo {year}
  {2013})}\BibitemShut {NoStop}%
\bibitem [{\citenamefont {Rubtsov}\ \emph {et~al.}(2008)\citenamefont
  {Rubtsov}, \citenamefont {Katsnelson},\ and\ \citenamefont
  {Lichtenstein}}]{PhysRevB.77.033101}%
  \BibitemOpen
  \bibfield  {author} {\bibinfo {author} {\bibfnamefont {A.~N.}\ \bibnamefont
  {Rubtsov}}, \bibinfo {author} {\bibfnamefont {M.~I.}\ \bibnamefont
  {Katsnelson}},\ and\ \bibinfo {author} {\bibfnamefont {A.~I.}\ \bibnamefont
  {Lichtenstein}},\ }\href {https://doi.org/10.1103/PhysRevB.77.033101}
  {\bibfield  {journal} {\bibinfo  {journal} {Phys. Rev. B}\ }\textbf {\bibinfo
  {volume} {77}},\ \bibinfo {pages} {033101} (\bibinfo {year}
  {2008})}\BibitemShut {NoStop}%
\bibitem [{\citenamefont {Werner}\ and\ \citenamefont
  {Millis}(2006)}]{PhysRevB.74.155107}%
  \BibitemOpen
  \bibfield  {author} {\bibinfo {author} {\bibfnamefont {P.}~\bibnamefont
  {Werner}}\ and\ \bibinfo {author} {\bibfnamefont {A.~J.}\ \bibnamefont
  {Millis}},\ }\href {https://doi.org/10.1103/PhysRevB.74.155107} {\bibfield
  {journal} {\bibinfo  {journal} {Phys. Rev. B}\ }\textbf {\bibinfo {volume}
  {74}},\ \bibinfo {pages} {155107} (\bibinfo {year} {2006})}\BibitemShut
  {NoStop}%
\bibitem [{\citenamefont {Li}\ and\ \citenamefont
  {Hanke}(2012)}]{PhysRevB.85.115103}%
  \BibitemOpen
  \bibfield  {author} {\bibinfo {author} {\bibfnamefont {G.}~\bibnamefont
  {Li}}\ and\ \bibinfo {author} {\bibfnamefont {W.}~\bibnamefont {Hanke}},\
  }\href {https://doi.org/10.1103/PhysRevB.85.115103} {\bibfield  {journal}
  {\bibinfo  {journal} {Phys. Rev. B}\ }\textbf {\bibinfo {volume} {85}},\
  \bibinfo {pages} {115103} (\bibinfo {year} {2012})}\BibitemShut {NoStop}%
\bibitem [{\citenamefont {Li}\ \emph {et~al.}(2008)\citenamefont {Li},
  \citenamefont {Lee},\ and\ \citenamefont {Monien}}]{PhysRevB.78.195105}%
  \BibitemOpen
  \bibfield  {author} {\bibinfo {author} {\bibfnamefont {G.}~\bibnamefont
  {Li}}, \bibinfo {author} {\bibfnamefont {H.}~\bibnamefont {Lee}},\ and\
  \bibinfo {author} {\bibfnamefont {H.}~\bibnamefont {Monien}},\ }\href
  {https://doi.org/10.1103/PhysRevB.78.195105} {\bibfield  {journal} {\bibinfo
  {journal} {Phys. Rev. B}\ }\textbf {\bibinfo {volume} {78}},\ \bibinfo
  {pages} {195105} (\bibinfo {year} {2008})}\BibitemShut {NoStop}%
\bibitem [{\citenamefont {Salpeter}\ and\ \citenamefont
  {Bethe}(1951)}]{PhysRev.84.1232}%
  \BibitemOpen
  \bibfield  {author} {\bibinfo {author} {\bibfnamefont {E.~E.}\ \bibnamefont
  {Salpeter}}\ and\ \bibinfo {author} {\bibfnamefont {H.~A.}\ \bibnamefont
  {Bethe}},\ }\href {https://doi.org/10.1103/PhysRev.84.1232} {\bibfield
  {journal} {\bibinfo  {journal} {Phys. Rev.}\ }\textbf {\bibinfo {volume}
  {84}},\ \bibinfo {pages} {1232} (\bibinfo {year} {1951})}\BibitemShut
  {NoStop}%
\end{thebibliography}%

\clearpage
\onecolumngrid
\appendix
\renewcommand\thefigure{S\arabic{figure}}    
\setcounter{figure}{0}   

\begin{center}
{\large\it Supplementary Information}\\
\vspace{0.5cm}

{\large\bf Doniach phase diagram for Kondo lattice model on the square and triangular lattices}

\vspace{0.3cm}
Ruixiang Zhou,$^{1}$
Xuefeng Zhang, $^{1}$
and Gang Li$^{1,2,\dagger}$ \\
\vspace{0.2cm}
${}^1${\small\it School of Physical Science and Technology, ShanghaiTech University, Shanghai 201210, China}\\
${}^2${\small\it Laboratory for Topological Physics, ShanghaiTech University, Shanghai 201210, China}\\

\end{center}

\vspace{0.2cm}
In this work, we study a model effectively describing the low-energy excitations of $f$-electrons. 
$f$-electrons are microscopically confined in spatially localized orbitals, rendering them a very small bandwidth in the electronic structure,
which yields a significantly enhanced electronic correlation $U$. 
The presence of spatially more extended wavefunctions, such as the $s$ and $p$ orbitals, can hybridize with the localized $f$-electrons with coupling strength $V$. 
In the limit of $U\rightarrow\infty$, a half-filled $f$-band will split into the lower and upper Hubbard bands. 
The charge fluctuations are completely suppressed leaving only the spin degrees of freedom active in low energy. 
The effective low-energy description of this process is given by the Kondo lattice model (KLM)~\cite{coleman2015introduction}.
\begin{equation}\label{KLM}
H=\sum_{k\sigma}\epsilon_{k}c_{k\sigma}^{\dagger}c_{k\sigma}+J \sum_i \vec{S}_i^f \cdot \vec{s}_i^c\;,
\end{equation}
where $J$ is the Kondo coupling derived as $J=4V^{2}/U$ from the associated periodic Anderson model (PAM) at half-filling. 

We want to provide a detailed comparison of the KLM on the square and triangular lattices focusing on the geometric influence on the magnetic long-range order and Kondo singlet. 
This requires an efficient evaluation of the metal-insulator transition (MIT) and magnetic susceptibility with high momentum resolution. 
Strong- or weak-coupling perturbation methods are not adequate to address the full parameter space. 
Methods restricted to a small system may suffer the finite-size effect. 
Geometrical frustration in the triangular lattice brings the ``minus-sign" problem to some approaches based on quantum Monte Carlo limiting their application to some electron occupancies.   
To this end, we use the dynamical mean-field theory (DMFT)~\cite{georges1996dynamical, burdin2002heavy, otsuki2009evolution, peters2013charge} and its non-local extension~\cite{martin2010fermi, martin2008evolution, maier2005quantum, PhysRevB.102.155126, li2013kondo, PhysRevLett.115.036404} in the framework of dual-fermion (DF) expansion~\cite{PhysRevB.77.033101, PhysRevB.79.045133}.
DMFT is a numerically exact approach applying to both small and large Kondo coupling regimes but it neglects all the non-local fluctuations of the conduction electrons. 
The local correlations on the magnetic impurity site are maintained. 
However, the RKKY interaction, as a non-local exchange interaction, is approximately treated in DMFT. 
The DF approach, on the other hand, as a non-local extension of DMFT,  inherits all the advantages of DMFT but can provide a better description of the non-local RKKY interactions. 
The application of these two approaches is expected to give a systematic understanding of the KLM on the square and triangular lattices at arbitrary electron filling.
We present the details of the two methodologies in the following.

\section{I. DMFT and DF methodology}\label{Append}
In this work, we used the DMFT to solve the KLM model on the two lattices and further corrected its local approximation by applying the DF scheme. 
In the following, we briefly summarize the methodology. 
\subsection{a. DMFT}
In the DMFT, we solve the local part of the Hamiltonian with exact diagonalization to yield the eigenvalues and eigenstates, which are necessary for evaluating the trace in each Monte Carlo configuration. 
\begin{equation}\label{H_local}
H_{loc}^{i} = -\mu\sum_{\sigma}c_{i\sigma}^{\dagger}c_{i\sigma} +
J\vec{S}_{i}^{f}\cdot\vec{s}_{i}^{c}.
\end{equation}
To diagonalize the local Hamiltonian in Eq.~(\ref{H_local}), we employ the two conserved quantum numbers $(N, S_{z}^{tot})$, i.e., $[H_{loc}, N] = [H_{loc}, S_{z}^{tot}]=0$, and block diagonalize the $H_{loc}^{i}$ to reduce the dimension of the subspaces in the latter Monte Carlo calculation.  
As shown in Tab.~\ref{Eigen}, the maximal dimension of the subspace is 2. The trace of each Monte Carlo configuration can be efficiently evaluated by direct matrix product. 
\begin{table}[htbp]
  \caption{Eigenvalues and eigenstates of the local Hamiltonian
    Eq.~(\ref{H_local}) expressed in the particle-number basis. The Hilbert
    space is decoupled into different blocks with respect to 
    the total particle number $N$ and the $z$-component of the total spin
    operator $S_{z}^{tot}$. The first arrow in each eigenstate represents the
    $c$-electron spin, while the second arrow represents the spin of
    the local magnetic impurity.} 
  \label{Eigen}
  \centering
  \begin{tabular}{|c|c|c|c|}
\hline
\hline
    Energy & Eigenstates & Particle Number & $S_{z}^{tot}$\\ 
\hline
    0    &   $|1\rangle=|0, \downarrow\rangle$  & 0 & -1/2 \\ 
\hline
    0    &   $|2\rangle=|0, \uparrow\rangle$  & 0 & 1/2 \\ 
\hline
$J/4-\mu$   & $|3\rangle=|\downarrow, \downarrow\rangle$  & 1 & -1 \\ 
\hline
$-3J/4-\mu$ & $|4\rangle=\frac{1}{\sqrt{2}}(|\uparrow,
\downarrow\rangle-|\downarrow,\uparrow\rangle)$  &\multirow{2}{*}{1} &
\multirow{2}{*}{0} \\  
\cline{1-2}
$J/4-\mu$ & $|5\rangle=\frac{1}{\sqrt{2}}(|\uparrow,
\downarrow\rangle+|\downarrow,\uparrow\rangle)$  &  &  \\      
\hline
$J/4-\mu$   & $|6\rangle=|\uparrow, \uparrow\rangle$  & 1 & 1 \\ 
\hline
$-2\mu$   & $|7\rangle=|\uparrow\downarrow, \downarrow\rangle$  & 2 & -1/2 \\ 
\hline
$-2\mu$   & $|8\rangle=|\uparrow\downarrow, \uparrow\rangle$  & 2 & 1/2 \\ 
\hline
  \end{tabular}
\end{table}

The kinetic energy, in the DMFT, maps to the hybridization function in the effective Anderson-type model. 
The hybridization function $\Delta(i\nu_{n})$ is dynamic. 
The DMFT equation is more conveniently expressed as an action
\begin{eqnarray}\label{dmft}
S_{imp} = H_{loc} + 
\sum_{\sigma}c_{\sigma,\nu_{n}}^{*}\Delta(i\nu_{n})c_{\sigma,\nu_{n}}  
\end{eqnarray}
We employed the continuous-time quantum Monte Carlo in the hybridization expansion scheme (CT-HYB) to solve the impurity action (\ref{dmft})~\cite{PhysRevB.74.155107}. 
With the prior eigenvalues and eigenstates calculated in Tab.~\ref{Eigen}, CT-HYB solves Eq.~(\ref{dmft}) by iteratively expanding the hybridization in the corresponding imaginary-time space $c_{\sigma}^{*}(\tau)\Delta(\tau - \tau^{\prime})c_{\sigma}(\tau^{\prime})$. 
The fermionic statistics of the $c$-fermions yields a determinant matrix of $\Delta$ over all combinations of time differences and a list of operators residing at different imaginary times. 
The dimension of the determinant matrix varies in the Monte Carlo update to explore all configurations in satisfying ergodicity. 
The trace of the $c$-operator product is directly evaluated with the help of an a priori compatibility check between different subspaces connected by the operators.  
The lowest temperature we measured in this work is $T/t = 0.01$, which is sufficient to illustrate our conclusion. 
Even lower temperatures can be also reached with the current algorithm. 

 \subsection{b. Dual Fermion approach}
To go beyond the local approximation of the DMFT, we consider the
non-local corrections generated by the local two-particle vertices,
through the dual-fermion approach~\cite{PhysRevB.77.033101,
  PhysRevB.79.045133}. 
The basic idea of the DF is to treat the difference between the impurity action in Eq.~(\ref{dmft}) and the following exact action as a small parameter.
\begin{eqnarray}\label{lattice}
S=-\sum_{k,\sigma}c^{*}_{\alpha}[i\nu_{n}+\mu
  -\epsilon_{k}]c_{\alpha}+J\sum_{i}\vec{S}^{f}\cdot\vec{s}^{c}\;,
\end{eqnarray}
where $\alpha$ denote $\alpha=(k,\omega_{n},\sigma)$. As the interaction is local, one can express the above lattice action as a collection of DMFT impurity actions, 
\begin{equation}
S=\sum_{i}S_{imp}^{i} +
\sum_{\alpha}c^{*}_{\alpha}
    [\epsilon_{k}-\Delta(i\nu_{n})] c_{\alpha}\;,
\end{equation}
where $S_{imp}$ is given by Eq.~(\ref{dmft}).
As stated above, if we take $\epsilon_{k}-\Delta(i\omega_{n})$ as an
expansion parameter (no matter it is large or small), we can formally
expand this action around $S_{imp}$, which gives rise to a non-local correction to the DMFT solution order by order. 

DF method introduces a new variable to achieve such an expansion. 
The last term in the above expression can be rewritten in an equivalent form 
with a new set of variables, {\it e.g.} $\{d\}$.
\begin{eqnarray}
e^{-\sum_{\alpha}c_{\alpha}^{\dagger}A_{\alpha}c_{\alpha}}
=\det^{-1}A
\int{\cal D}[d,d^{*}]e^{(-c^{*}_{\alpha}d_{\alpha}+h.c.) 
  - d^{*}_{\alpha}A_{\alpha}d_{\alpha}}\;.
\end{eqnarray}
The integration now depends on both the conduction, {\it
  i.e.} $\{c\}$, and the dual variables, {\it i.e.} $\{d\}$. 
Here $A$ denotes matrix
$\widehat{A}_{k\omega_{n},\sigma}$, with element
$A_{k\omega_{n},\sigma}=[\Delta(i\nu_{n})-\epsilon_{k}]^{-1}$. 
The partition function of the lattice problem defined in
Eq.~(\ref{lattice}) now becomes dependent on all three variables,
\begin{eqnarray}\label{new-action}
{\cal Z} &=&\det^{-1}A\int 
{\cal D}[c,c^{*};f,f^{*}]e^{-\sum_{i}S_{imp}^{i}}\int {\cal
  D}[d,d^{*}]e^{-\sum_{\alpha}[c_{\alpha}^{*}d_{\alpha} + d_{\alpha}^{*}c_{\alpha} 
  +d_{\alpha}^{*}A_{\alpha}d_{\alpha}]}\phantom{.} 
\end{eqnarray}

We should note here, there is no approximation involved in all above
derivations, i.e., Eq.~(\ref{new-action}) is an equivalent expression of the lattice
action.  
Thus, any quantity of interests can be equally evaluated through
Eq.~(\ref{new-action}) and Eq.~(\ref{lattice}).
For example, one can calculate the single-particle Green's function
from both actions (by taking $\epsilon_{k}$ as source term and
  differentiate both actions over it), which leads to
\begin{equation}\label{G-fd}
G_{k\nu_{n},\sigma}=[\Delta(i\nu_{n})-\epsilon_{k}]^{-2}G^{dual}_{k\nu_{n},\sigma}
+ [\Delta(i\nu_{n})-\epsilon_{k}]^{-1}, 
\end{equation}
where $G^{dual}_{k\nu_{n},\sigma}=-\langle
d_{k\nu_{n},\sigma}d^{*}_{k\nu_{n},\sigma}\rangle$
denotes the single-particle Green's function of the dual variables. 

It becomes transparent now that introducing the dual variable changes
the calculation of the lattice Green's function $G_{k\nu_{n},\sigma}$ to
that of $G^{dual}_{k\nu_{n},\sigma}$.
Concerning the complexity of solving the Kondo Hamiltonian, it
seems that nothing is achieved in the transformation in Eqs.~(\ref{lattice}
- \ref{G-fd}), as $G^{dual}_{k\nu_{n},\sigma}$ is not known and its
calculation can be equally complicated.
However, as one will see below because the expansion is
around the DMFT solution, $G_{k\nu_{n},\sigma}^{dual}$ can be
calculated perturbatively, which is much simpler than the
calculation of $G_{k\nu_{n},\sigma}$ directly.

As for the Kondo problem we study in this work, the evaluation of
$G^{dual}_{k\omega_{n},\sigma}$ formally requires an action that
depends only on $\{d\}$ and $\{f\}$, which can
be obtained by integrating $\{c\}$ out of Eq.~(\ref{new-action}).
$\{c\}$ and $\{d\}$ are separated in Eq.~(\ref{new-action}) except for $ 
(c_{k\nu_{n},\sigma}^{*}d_{k\nu_{n},\sigma}+h.c.)$. 
Expanding the partition function in Eq.~(\ref{new-action}) over
this mixed term, and neglecting any term in which $c$ and $c^{*}$ are
not paired (according to the Grassmann algebra) leads to 
\begin{eqnarray}\label{dual-Z}
{\cal Z} &=&\det^{-1}A\int
     {\cal D}[d, d^{*}] 
     \exp(-\sum_{\alpha}d^{*}_{\alpha}A_{\alpha}d_{\alpha})\times{\cal Z}_{imp}\int{\cal
       D}[c,c^{*};f,f^{*}]e^{-\sum_{i}S_{imp}^{i}[c,c^{*};f,f^{*}]}\times\nonumber\\
     &&\times[1+\sum_{\alpha_{1}\alpha_{2}}d^{*}_{\alpha_{1}}\langle
       c_{\alpha_{1}}c_{\alpha_{2}}^{*}
       \rangle_{imp}d_{\alpha_{2}} 
     +\frac{1}{4}\sum_{\alpha_{1}\alpha_{2}\alpha_{3}\alpha_{4}}\langle
       c_{\alpha_{1}}c_{\alpha_{2}}^{*}c_{\alpha_{3}}c_{\alpha_{4}}^{*}\rangle_{imp} 
     d_{\alpha_{1}}^{*}d_{\alpha_{2}}d_{\alpha_{3}}^{*}d_{\alpha_{4}}
     + \mbox{higher orders}] \nonumber\\
     &=&{\cal Z}_{imp}\det^{-1}A\int{\cal D}[d, d^{*}] 
     \exp(-\sum_{\alpha}d^{*}_{\alpha}(A_{\alpha}+g_{\alpha})d_{\alpha} - V[d,d^{*}]),
\end{eqnarray}
where $g_{\alpha}=-\langle c_{\alpha}c_{\alpha}^{*}\rangle_{imp}$ and
$\chi_{1234}=\langle
c_{\alpha_{1}}c_{\alpha_{2}}^{*}c_{\alpha_{3}}c_{\alpha_{4}}^{*}\rangle_{imp}$
are the impurity one- and two-particle Green's functions.
A complete separation of variables $\{d\}$ with the local degrees of
freedom $\{c\}$ and $\{f\}$ is achieved in this equation.
In the last step, we have brought all expansion terms back to the exponential
function and group all terms of order higher than 1 to $V[d,d^{*}]$. 
The first term in $V[d, d^{*}]$ is 
$\frac{1}{4}\gamma_{1234}d_{\alpha_{1}}^{*}d_{\alpha_{2}}d_{\alpha_{3}}^{*}d_{\alpha_{4}}$,
with $\gamma_{1234}$ the reducible part of $\chi_{1234}$ (, the bubble
the term is subtracted due to the exponential form of
$\sum_{\alpha_{1}\alpha_{2}}d^{*}_{\alpha_{1}}\langle c_{\alpha_{1}}c_{\alpha_{2}}^{*} 
       \rangle_{imp}d_{\alpha_{2}}$). 

$g_{\alpha}$ and $\gamma_{1234}$ are given as the solutions of the DMFT
calculations. 
As a technical remark, we note that, with the CT-HYB as the impurity
solver, the higher-frequency part of the Matsubara self-energy
the function contains larger statistical error than the lower-frequency
part.  
In contrast, the sampling of $g_{\alpha}$ and $\gamma_{1234}$ is more
stable and less affected by the statical noises.
And due to the subtraction of the bubble contribution from
$\chi_{1234}$, $\gamma_{1234}$ can be reliably sampled directly in the
Matsubara frequency space~\cite{PhysRevB.85.115103}.

Eq.~(\ref{dual-Z}) is exactly same as the original lattice partition in Eq.~(\ref{lattice}), no approximation is introduced 
either in changing variables or in the expansion of
the mixed term as long as we carefully collect every term in
$V[d,d^{*}]$.   
However, as $V[d,d^{*}]$ are vertex functions and difficult to deal with, we are particularly
limited to a few lower-order terms of it.
We only considered the two-particle reducible vertex in $V[d,d^{*}]$.   
It turns out to be sufficient, at least at a qualitative level, for constructing the non-local corrections to the DMFT solution in most cases.

With the action of the dual variables,
\begin{equation}
S[d,d^{*}] =
\sum_{\alpha}d_{\alpha}^{*}G_{0,\alpha}^{dual,-1}d_{\alpha} + V[d^{*},d],
\end{equation}
one can perturbatively determine $G_{k\nu_{n},\sigma}^{dual}$ from
$G_{0,k\nu_{n},\sigma}^{dual}=1/[(\Delta(i\nu_{n})-\epsilon_{k})^{-1}
  + g_{\nu_{n},\sigma}]$
and $V[d, d^{*}]$. 
In terms of the two-particle reducible vertex, the self-energy
the function of the dual variables can be calculated, {\it e.g.} from the
first two diagrams~\cite{PhysRevB.77.033101, PhysRevB.79.045133,PhysRevB.78.195105} in an expansion of $V[d, d^{*}]$, as 
\begin{eqnarray}
\Sigma^{dual}_{\alpha_{1}}&=&[\frac{T}{N}\sum_{2,3,4}G^{dual}_{\alpha_{3}}\gamma_{1234}
(\delta_{\alpha_{2};\alpha_{3}}-\delta_{\alpha_{1};\alpha_{2}})-\frac{T^{2}}{2N^{2}}\sum_{2,3,4}G_{\alpha_{2}}^{dual}
G_{\alpha_{3}}^{dual}
G_{\alpha_{4}}^{dual}\gamma_{1234}\gamma_{4321}]\delta_{\alpha_{1}+\alpha_{3};\alpha_{2}+\alpha_{4}}.  
\end{eqnarray}
$G_{\alpha}^{dual}$ and $\Sigma_{\alpha}^{dual}$ are
related by Dyson equation $G_{\alpha}^{dual,
  -1}=G_{0,\alpha}^{dual, -1} -\Sigma_{\alpha}^{dual}$.
Through Eq.~(\ref{G-fd}), the frequency and
momentum dependent Green's function $G_{k\nu_{n},\sigma}$ are obtained. 
\section{II. Additional results}
\subsection{a. Magnetic phase boundary}

In the doniach phase diagram, we obtained the AFM long-range magnetic phase from the instability of the paramagnetic solution. We restricted our calculations in the paramagnetic phase by eliminating the spin polarization in the self-energy and single-particle Green's function, while the spin susceptibility, as the fluctuations of the magnetic moments, memorizes the spin polarization and the magnetic instability.  
Moreover, the divergence of the magnetic susceptibility further tells the spin symmetry in the broken phase, which is an unbiased way of understanding spontaneous spin symmetry breaking. 
Another way of study is to compare the free energy of states with different magnetic symmetry-breaking modes. 
The one with the lowest free energy is taken as the candidate of the spin-polarized state.   
While such a strategy requires the exploration of all kinds of symmetry broken phases, which are usually impractical to exhaust or difficult to implement in practice.  
In this sense, the evaluation of spin susceptibility with high momentum resolution has the obvious advantage as it does not predefine any magnetic symmetry-breaking but directly obtains it from the divergence of the spin susceptibility. 

For interacting systems, spin susceptibility consists of the contributions from the bare bubble diagram and the two-particle (2P) vertex function $\Gamma^{k,k^{\prime};q}_{m}$.
\begin{eqnarray}\label{Eq:susceptibility}
\chi(\vec{q}, i\Omega_{m}) =  \chi_{0}(\vec{q}, i\Omega_{m})
+ \frac{T^{2}}{N^{2}}\sum_{k,k^{\prime}}G^{\prime}(k)G^{\prime}(k+q)\Gamma_{m}^{k,k^{\prime};q}G^{\prime}(k^{\prime}+q)G^{\prime}(k^{\prime})\;,
\end{eqnarray} 
where $\chi_{0}(\vec{q}, i\Omega_{m})$ is the convolution of two one-particle (1P) interacting Green's function, 
\begin{equation}
\chi_{0}(\vec{q}, i\Omega_{m}) = -\frac{T}{N}\sum_{k}G(k)G(k+q)\;.
\end{equation}
Here, $k=(\vec{k}, i\nu_{n})$ and $q=(\vec{q}, i\Omega_{m})$ are joint variables referring to both momentum and frequency. 
In the DMFT and DF formalisms, $G(k)$, $G^{\prime}(k)$, and $\Gamma_{m}^{k, k^{\prime};q}$ have different definitions.
In DMFT, $G(k)=G^{\prime}(k) = 1/(i\nu_{n}+\mu - \epsilon_{k} - \Sigma(i\nu_{n}))$ is the DMFT lattice Green's function with local self-energy. 
$\Gamma_{m}^{k,k^{\prime};q}=\Gamma_{m}^{\nu_{n},\nu_{n}^{\prime};q}$ is the 2P vertex function calculated from the local one $\gamma_{m}^{\nu_{n},\nu_{n}^{\prime};\Omega_{m}}$ via the Bethe-Salpeter equation (BSE)~\cite{PhysRev.84.1232}.
In the DF method, $G(k)=1/(i\nu_{n} + \mu - \epsilon_k - \Sigma(k))$ where self-energy is fully momentum-dependent. 
$\Gamma_{m}^{k,k^{\prime};q}$ is also calculated through BSE but the corresponding bubble term is composed of the DF Green's function $G^{d}(k)$ instead of $G(k)$. 
It is, thus, the dressed 2P full vertex function of the dual fermions.
To connect it with the conduction electrons in the susceptibility, $G^{\prime}(k) = G^{d}(k)/[g_{\nu}(\Delta_{\nu}-\epsilon_{k})]$ has to be defined in Eq.~(\ref{Eq:susceptibility}). 

As can be seen from the above, the non-local fluctuations come into the DF susceptibility in both the 1P Green's function and 2P vertex function.
The resulting modification yields a significant suppression of the spin susceptibility, especially in the intermediate $J$ regime. 
It is easy to understand why DMFT becomes a good approximation in the weak and strong coupling regimes concerning the magnetic correlations. 
The non-local corrections to the spin susceptibility are mainly through the momentum-dependent 1P self-energy $\Sigma(k)$, which is small when $J$ is small. 
For large $J$, due to the formation of the Kondo singlet, the RKKY coupling is no longer active. Both the DMFT and DF lead to the absence of long-range AFM order, i.e., they are no longer different in spin susceptibility. 
Fluctuations become relevant in the phase boundary as expected, which pronouncedly shrinks the AFM phase on the square lattice, as shown in Fig.~\ref{doniach_diagram} of the main text.  

\subsection{b. Self-energy and non-local corrections}
\label{Sigma}

\begin{figure}[htbp]
\centering
\includegraphics[width=0.8\linewidth]{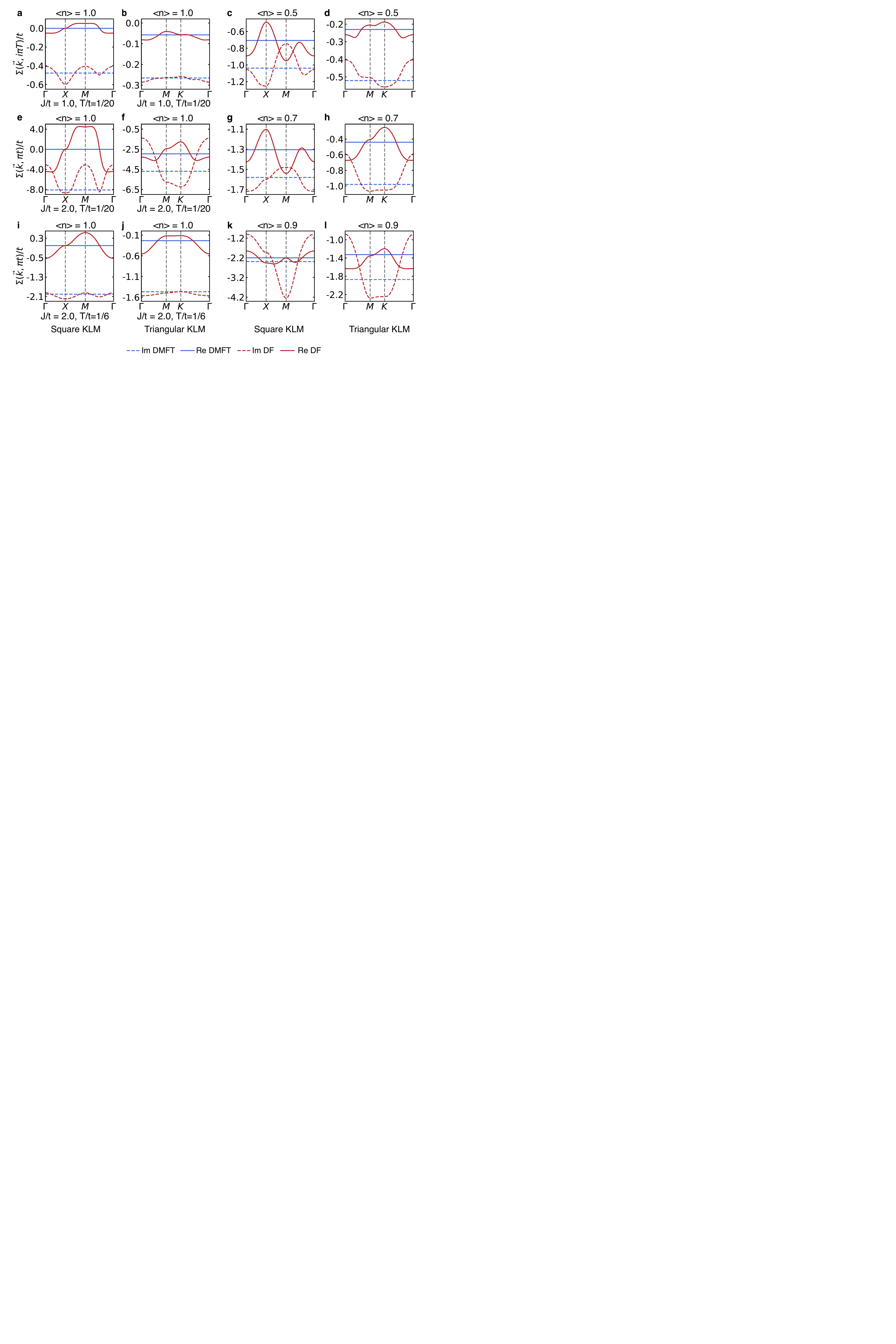}
\caption{\textbf{(Color Online) Non-local correction to the self-energy.} The first two columns display the self-energy of the half-filled KLM on the square and triangular lattice with the corresponding parameters shown under each subplot. The blue and red lines are the solutions from the DMFT and DF calculations, where the solid/dashed line corresponds to the real/imaginary part of the self-energy. The third and the fourth columns show the comparison with the same parameter $J/t = 1.8, T/t = 1/20$ but at different hole doping levels. }
\label{Fig_Sigma}
\end{figure}

Both the doniach phase diagram Fig.~\ref{doniach_diagram} and the MIT phase diagram Fig.~\ref{Fig_MIT} display visible non-local corrections to the DMFT solutions, which pronouncedly shrink the AFM phase on the square lattices and reduce the critical values of $J_{c}$ for the MIT. 
To be more clear,  we inspect the non-local corrections by comparing the self-energy from the DMFT and DF calculations at half-filling. 
In Fig.~\ref{Fig_Sigma} we show the momentum-dependent self-energy $\Sigma(\vec{k}, i\pi T)$ at the lowest Matsubara frequency along the conventional high-symmetry $k$-path for the square and triangular lattice, respectively. 
The first and the third columns display the calculations for the square lattice. The second and the fourth columns are the same calculations but for the triangular lattice. 
The DMFT and DF solutions of $\Sigma(\vec{k}, i\pi T)$ correspond to the blue and red lines. 
The solid and dashed lines denote the real and imaginary parts of the self-energy, respectively. 

DMFT is a local approximation,  $\Sigma(\vec{k}, i\pi T)$ is a constant in the entire BZ. 
In all plots, the DMFT solutions are straight lines along the $k$-path indicating the complete independence of the momentum in this method. 
On the other hand, after partially including non-local fluctuations, the DF solutions yield a dispersive self-energy with variation depending on the parameters. 
The analysis of the self-energy dispersion in momentum space is extremely useful in justifying the validity of the DMFT approximation.  
Let us take the comparison in Fig.~\ref{Fig_Sigma}\textbf{a} and Fig.~\ref{Fig_Sigma}\textbf{b} as an example. The parameter is $J/t = 1.0$, $T/t = 1/20$, at which both the DMFT and DF solutions are metallic according to Fig.~\ref{Fig_MIT}. 
The DF self-energy dispersion shown in red curves fluctuates around the DMFT constant values. By increasing the coupling to $J/t = 2.0$ shown in \textbf{e} and \textbf{f}, the DF self-energy becomes much more dispersive.
Furthermore, we notice that similar to the case with $J/t=1.0$, the real part of the DF self-energy is more or less around the DMFT solutions, i.e., the local self-energy in the DF calculations is close to the DMFT self-energy $\frac{1}{N_{k}}\sum_{k}\mbox{Re}\Sigma^{\mbox{DF}}(\vec{k}, i\pi T)\sim \mbox{Re}\Sigma^{\mbox{DMFT}}(i\pi T)$. 
However, the imaginary part of the DF self-energy, after coarse-graining over momentum, deviates from that of the DMFT self-energy (see Fig.~\ref{Fig_Sigma}\textbf{e}), highlighting the substantial non-local corrections to the DMFT. 
Such non-local corrections are more pronounced at the low temperature and stronger coupling $J$ regime, indicating the insufficiency of the DMFT local approximation for these parameters.  
If one compares the non-local corrections on the square and triangular lattice in Fig.~\ref{Fig_Sigma}, it is easy to find that they are more pronounced on the square lattice but are somewhat suppressed on the triangular lattice due to the geometrical frustration. 
While the amplitude of the non-local corrections on the triangular lattice is still very large, which cannot be simply neglected. 
Such non-local corrections bring significant corrections to the MIT phase boundary in Fig.~\ref{Fig_MIT}\textbf{c, d}, which reminds us that, even on a geometrically frustrated lattice, the non-local fluctuations can still be very strong leading to significant corrections to the macroscopic response. 
Plots shown in Fig.~\ref{Fig_Sigma}\textbf{i} and \textbf{j} correspond to calculations at a higher temperature. The self-energy dispersion becomes less obvious than those at lower temperatures. 
Furthermore, by comparing the self-energy at different doping levels, we observe that, with the increase of the hole-doping, the self-energy becomes less dispersive indicating the DMFT local approximation becomes more justified in the doped case. 
Thus, we conclude that the non-local corrections to the self-energy is more pronounced at half-filling and lower temperature regime, where the DMFT is less justified. 

\subsection{c. $T_{K}$ from uniform susceptibility}
\label{UniSus}

To estimate the Kondo temperature in Fig.~\ref{doniach_diagram}, we calculate the impurity uniform susceptibility, 
\begin{equation}
\chi^{imp}(T) = \int_{0}^{\beta}d \tau \chi^{imp}(\tau, T) = \int_{0}^{\beta}d\tau \langle S_{z}(\tau)S_{z}(0)\rangle\;,
\end{equation}
and normalize it with the static susceptibility $C_{Z}=\langle S_{z}^{2}\rangle$. 
With CT-HYB being the impurity solver for the DMFT, both $\chi^{imp}(T)$ and $C_{Z}$ can be efficiently measured to very high precision. 
The Kondo temperature $T_{K}$ can be estimated from $T_{K}\sim C_{Z}/\chi^{imp}$ at sufficiently low temperatures. 

\begin{figure}[htbp]
	\centering
	\includegraphics[width=0.8\linewidth]{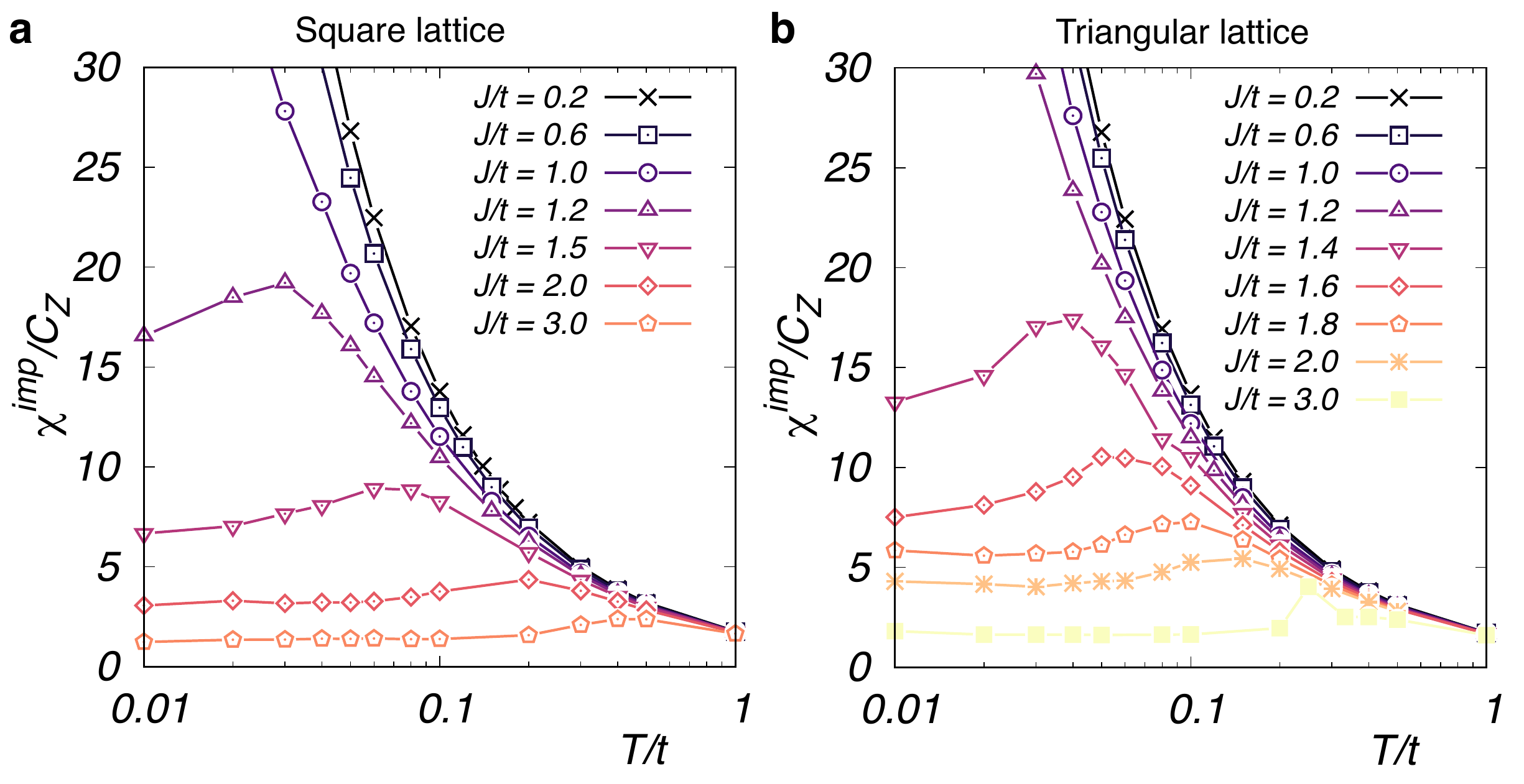}
	\caption{
\textbf{(Color Online) The impurity uniform susceptibility as a function of temperature for different coupling strengths.}
\textbf{a} and \textbf{b} are for the half-filled square and triangular lattice, respectively.
\label{Fig_UniSus}
}
\end{figure}

Figure~\ref{Fig_UniSus} display the $\chi^{imp}/C_{Z}$ as a function of temperature. 
We show the corresponding results at a few different coupling strengths in each plot. 
Overall, the uniform susceptibility shows a Curie-Weiss paramagnetic scaling of $1/T$ at high temperatures.
With the decrease in temperature, it first bends down and then gradually saturates to a constant value showing a Pauli paramagnetic behavior. 
Comparing the square and triangular lattice, we found that, for the same coupling strength $J$, the characteristic temperature $T^{*}$, at where the uniform susceptibility starts to bend, is slightly higher on the triangular lattice.  
Moreover, the saturation values at the lowest temperature, i.e., $T/t = 0.01$ in this work, are slightly higher than those on the square lattice. 
Their difference becomes smaller with the increase of $J/t$.
As a consequence, the determined Kondo temperatures in the doniach phase diagram Fig.~\ref{doniach_diagram} for the square and triangular lattices display overall similar scaling behavior. 
The inverse of the uniform susceptibility is slightly smaller on the triangular lattice for the same Kondo coupling, indicating that the triangular lattice has a smaller Kondo temperature as compared to the square lattice. 

\subsection{d. Spectral function and the Kondo screening}
\begin{figure}[htbp]
\centering
\includegraphics[width=0.8\linewidth]{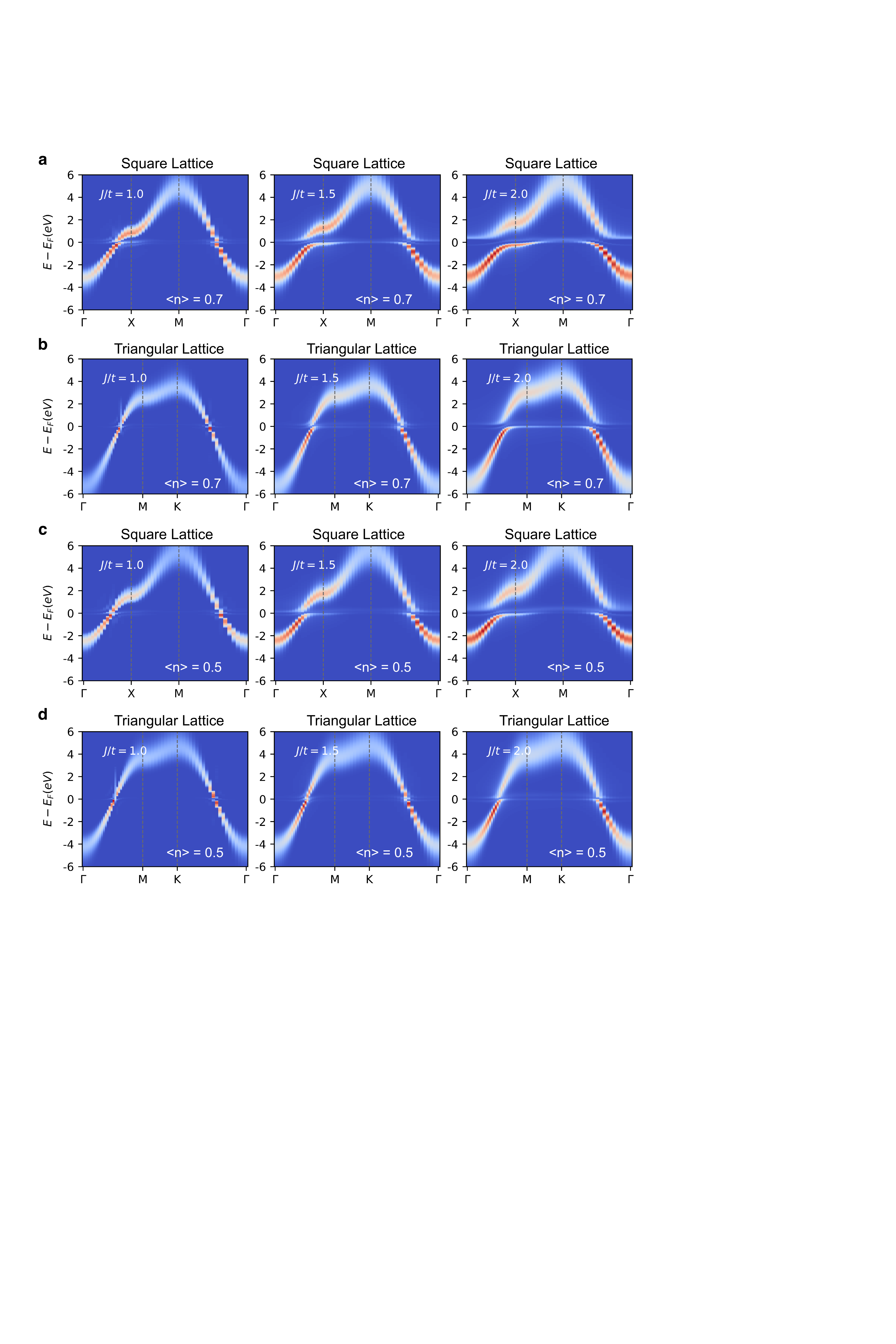}
\caption{\textbf{(Color Online) The spectral function comparison for the hole-doped KLM on the square and triangular lattice.}
The comparison is shown at two doping levels. \textbf{a} and \textbf{b} display the comparison at $\langle n\rangle = 0.7$ and $T/t = 0.04$ for three different coupling strengths $J/t=1.0, 1.5$, and $2.0$ as indicated in each subplot. 
\textbf{c, d} are same as \textbf{a, b} but for the electron density $\langle n\rangle = 0.5$. 
\label{Fig_spectra}
}
\end{figure}

Doping with holes, the KLM on both lattices becomes metallic. 
The Kondo screening and the RKKY coupling both behave differently as compared to the half-filled case. 
In the main text, we have confirmed that the Kondo screening exists even inside the magnetically ordered states, which leaves a clear signature of the shadow bands of the localized $f$-electrons in the conduction spectra. 
The shadow bands are not simply the folded bands from the magnetic BZ, but are the same as the bands observed in the Kondo regime. 
Doping with holes, by inspecting these shadow bands, we are able to monitor the Kondo screening as a function of the hole doping. 
In Fig.~\ref{Fig_spectra} we show the spectral function comparison for the doped KLM on the square and triangular lattice at two different doping levels, i.e., $\langle n\rangle=0.7$ and $0.5$. 
An overall similarity to the half-filled case is observed except for the absence of the charge gap in the spectra. 
The Fermi level shifts like a rigid band with hole doping on both square and triangular lattices. 

With the increase of hole doping, the intensity of the shadow bands gradually becomes weaker, indicating a suppressed Kondo screening with doping. 
Furthermore, if one compares the shadow bands on the two lattices at the same hole doping, one will notice that the Kondo screening strength is slightly stronger on the square lattice than on the triangular lattice, which is consistent with the estimation from the uniform susceptibility at half-filling. 
Geometrical frustration, thus, suppresses both the RKKY coupling and the Kondo coupling.

\end{document}